\documentclass[twocolumn]{aastex62}

\usepackage[utf8]{inputenc}
\usepackage{longtable}
\usepackage{hyperref}
\usepackage{textcomp}
\usepackage{tabularx}
\usepackage{wrapfig}
\usepackage{float}
\usepackage{here}
\usepackage{amsmath}
\usepackage{rotating}
\usepackage{blindtext}
\usepackage{color}
\usepackage[section]{placeins}
\usepackage{natbib}

\newcommand{\allstars}{327} 

\graphicspath{{./}{figures/}}

\shorttitle{STELLAR COMPANIONS TO ASTEROSEISMIC STARS}
\shortauthors{J. SCHONHUT-STASIK ET AL.}

\begin{document}

\title{ROBO-AO \textit{KEPLER} ASTEROSEISMIC SURVEY. II. DO STELLAR COMPANIONS INHIBIT STELLAR OSCILLATIONS?}

\correspondingauthor{Jessica Schonhut-Stasik}
\email{jstasik@hawaii.edu}

\author[0000-0002-1043-8853]{Jessica Schonhut-Stasik}
\affiliation{Institute for Astronomy, University of Hawai`i at M\={a}noa, Hilo, HI 96720-2700, USA}

\author[0000-0001-8832-4488]{Daniel Huber}
\affiliation{Institute for Astronomy, University of Hawai`i at M\={a}noa, Honolulu, HI 96822-1839, USA}

\author[0000-0002-1917-9157]{Christoph Baranec}
\affiliation{Institute for Astronomy, University of Hawai`i at M\={a}noa, Hilo, HI 96720-2700, USA}

\author[0000-0002-6731-9329]{Claire Lamman}
\affiliation{Department of Astrophysical and Planetary Sciences, University of Colorado Boulder, Boulder, CO 80309, USA}

\author[0000-0002-5082-6332]{Ma\"{i}ssa Salama}
\affiliation{Institute for Astronomy, University of Hawai`i at M\={a}noa, Hilo, HI 96720-2700, USA}

\author[0000-0003-0054-2953]{Rebecca Jensen-Clem}
\affiliation{Astronomy Department, University of California, Berkeley, CA 94720, USA}

\author[0000-0001-5060-8733]{Dmitry A. Duev}
\affiliation{Division of Physics, Mathematics, and Astronomy, California Institute of Technology, Pasadena, CA 91125, USA}

\author[0000-0002-0387-370X]{Reed Riddle}
\affiliation{Division of Physics, Mathematics, and Astronomy, California Institute of Technology, Pasadena, CA 91125, USA}

\author{S. R. Kulkarni}
\affiliation{Division of Physics, Mathematics, and Astronomy, California Institute of Technology, Pasadena, CA 91125, USA}

\author[0000-0001-9380-6457]{Nicholas M. Law}
\affiliation{Department of Physics and Astronomy, University of North Carolina at Chapel Hill, Chapel Hill, NC 27599-3255, USA}


\begin{abstract}
\label{sec:abstract}

The \textit{Kepler} space telescope observed over 15,000 stars for asteroseismic studies. Of these, 75\% of dwarfs (and 8\% of giants) were found to show anomalous behavior: such as suppressed oscillations (low amplitude) or no oscillations at all. The lack of solar-like oscillations may be a consequence of multiplicity, due to physical interactions with spectroscopic companions or due to the dilution of oscillation amplitudes from ``wide'' (AO detected; visual) or spectroscopic companions introducing contaminating flux. We present a search for stellar companions to \allstars{} of the \textit{Kepler} asteroseismic sample, which were expected to display solar-like oscillations. We used direct imaging with Robo-AO, which can resolve secondary sources at $\sim$0\farcs15, and followed up detected companions with Keck AO. Directly imaged companion systems with both separations of $\leq$ 0\farcs5 and amplitude dilutions $>$10\% all have anomalous primaries, suggesting these oscillation signals are diluted by a sufficient amount of excess flux. We also used the high-resolution spectrometer ESPaDOnS at CFHT to search for spectroscopic binaries. We find tentative evidence for a higher fraction of spectroscopic binaries with high radial velocity scatter in anomalous systems, which would be consistent with previous results suggesting that oscillations are suppressed by tidal interactions in close eclipsing binaries.

\end{abstract}



\section{Introduction}
\label{sec:introduction} 

Asteroseismology, the study of stellar oscillations, benefits from the wealth of data provided by the original \textit{Kepler} Mission \citep{Borucki2010}. By measuring brightness variations in \textit{Kepler} light curves, we can identify and study pulsations, which are then used to infer precise stellar parameters. 

\textit{Kepler} observed $\sim$2000 dwarfs and subgiants predicted to display solar-like oscillations\footnote{`Solar-like' refers to stellar oscillations excited by the same mechanism as the Sun: through turbulent convection in their outer layers.}, collecting over a month of short cadence data for each star. Surprisingly, detectable oscillations were only found in $\sim$500 of these stars \citep{Chaplin2011a}. 

To search for solar-like oscillations in red giant stars, \textit{Kepler} surveyed $\sim$20,000 giants using long cadence observations\footnote{Because the period of oscillation is longer in red giant stars (log(g)$<$3.5) this allows the use of long cadence observations.}. Giants, with their large pulsation amplitudes, should always exhibit oscillations above the \textit{Kepler} detection limit, however 1671 of these were classified as non-detections \citep{Hon2019}. As well as non-detections, some red giants show suppressed oscillations meaning a detection is made, but at a lower amplitude than expected. 

\begin{figure}
\centering
\includegraphics[width=0.5\textwidth]{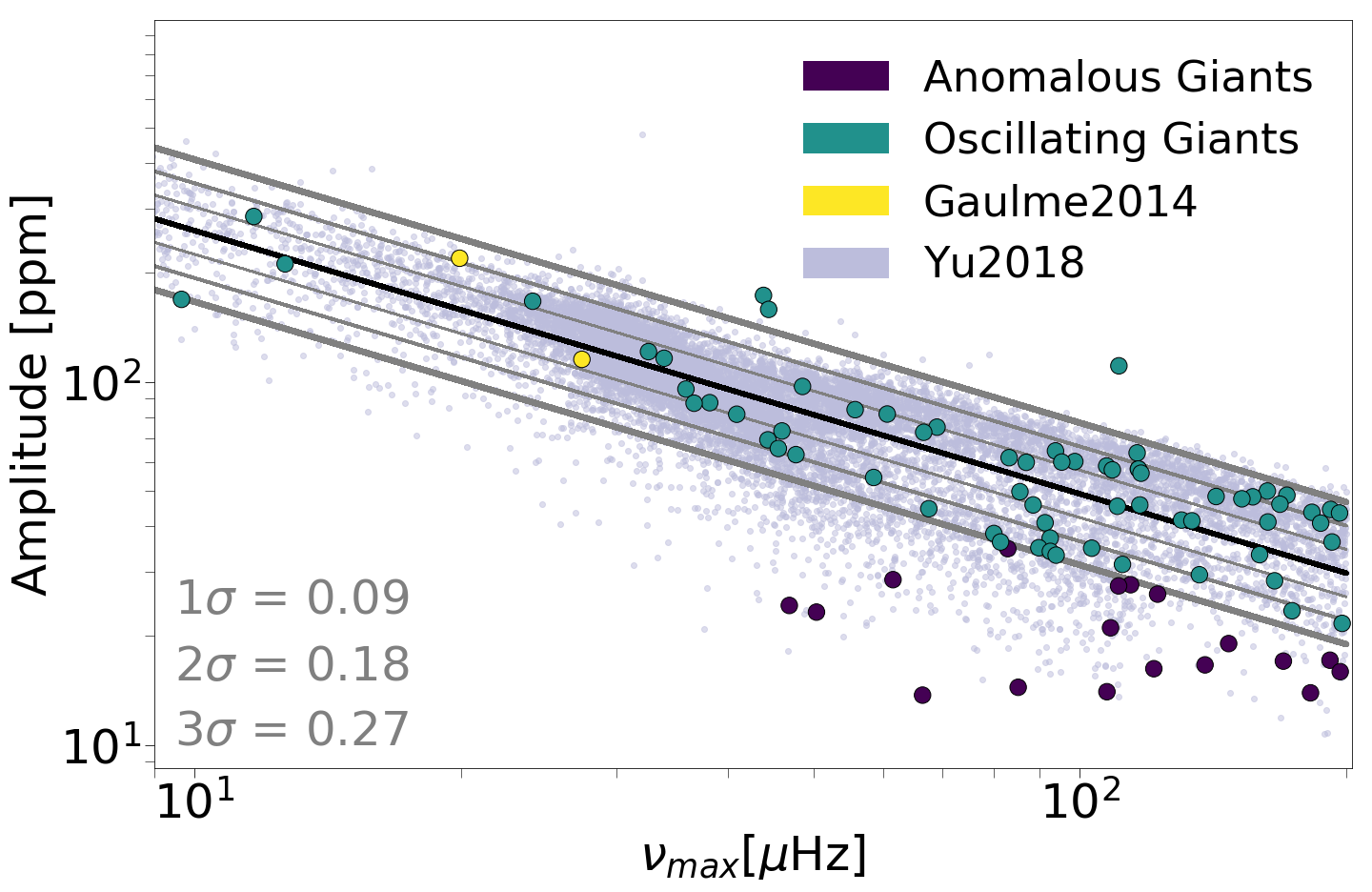}
\caption{The amplitude of oscillations versus frequency of maximum power for the red giants in our sample (colored circles) and the red giants in \cite{Yu2018} (light blue dots). The black line shows a linear fit whilst the gray lines correspond to 1, 2 and 3$\sigma$ limits (each of these limits is quantified in the bottom left of the plot). Green circles are oscillating stars whilst purple circles are anomalous. Yellow stars are part of the G14 sample. This sample is restricted to the \cite{Yu2018} selection criteria and therefore does not include the entirety of our red giant sample or the \cite{Gaulme2014} sample.}
\label{fig:numax_amp}
\end{figure}

\begin{table}[ht]
\centering
\caption{Robo-AO Sample Breakdown \label{tab:num_of_stars}}
\begin{tabular}{l|cc}
\hline
& \textbf{Dwarfs} & \textbf{Giants} \\
\hline
\textbf{Oscillating} & 100             & 99              \\
\textbf{Anomalous}   & 54              & 55              \\
\end{tabular}
\end{table}

This lack of oscillations could suggest a significant physical difference between stars sharing similar fundamental properties. Alternatively, inaccurate stellar properties could be used to mischaracterize a star as oscillating \citep{Chaplin2011b}. For example, the inferred oscillation amplitude of a star will be overestimated if the stellar type is based on an overestimated luminosity. In fact, the star may exhibit oscillation amplitudes which are too small to be observed by \textit{Kepler}.

A lack of oscillations could also be attributed to multiplicity; either via the dilution of amplitudes caused by contaminating flux \citep{schonhutstasik2017}, or by spectroscopic binaries\footnote{All spectroscopic binaries are physically associated.}, inhibiting oscillations through tidal interaction. Tidal interactions between stars are believed to increase magnetic activity and subsequently decreases the efficiency of the surface convection that drives oscillations, inducing amplitude suppression. \cite{Gaulme2014} (hereafter G14) demonstrated a link between amplitude suppression and close binaries using \textit{Kepler} observations of 19 red giant eclipsing binary systems. Fifteen of the red giants demonstrated solar-like oscillations, whilst there were oscillations detected in the remaining four. The stars with no mode detections exhibit shorter orbital periods (between 15 and 45 days). For individual modes, the relationship between oscillations and binarity has also been investigated. For example, it has been found that detached eclipsing binaries present p-dominated mixed-modes more often \citep{Themesl2017}.

It is plausible that systems can contain both a wide and spectroscopic companion, suggesting that multiple mechanisms can act simultaneously to suppress amplitudes. These systems can occur frequently. \cite{Tokovinin2006} found a 96\% likelihood that a solar-type spectroscopic binary system (with an orbital period of $<$3 days) will also contain a tertiary companion.

Despite the discovery of these links between oscillations and multiplicity, there have been no large-scale statistical studies on the effects of multiplicity on oscillation formation and detection. 

In this work we investigate the effect of multiplicity on stellar oscillations, through a large combined imaging and spectroscopic campaign. 
We identify wide companions which may cause amplitude dilution by observing 327 \textit{Kepler} asteroseismic stars using Robo-AO. We search for spectroscopic companions to stars that may be causing tidal interference using ESPaDOnS at the Canada France Hawai'i Telescope (CFHT). ESPaDOnS performed multi-epoch, high-resolution spectroscopy for a sub-sample of 34 targets containing both single stars and wide binaries. Our imaging sample also contains the 19 red giant eclipsing binaries from G14. Imaging these stars will determine whether a wide companion is also present in their system, building on the findings of \cite{Tokovinin2006}.

\begin{figure*}[ht]
\centering
\includegraphics[width=\textwidth]{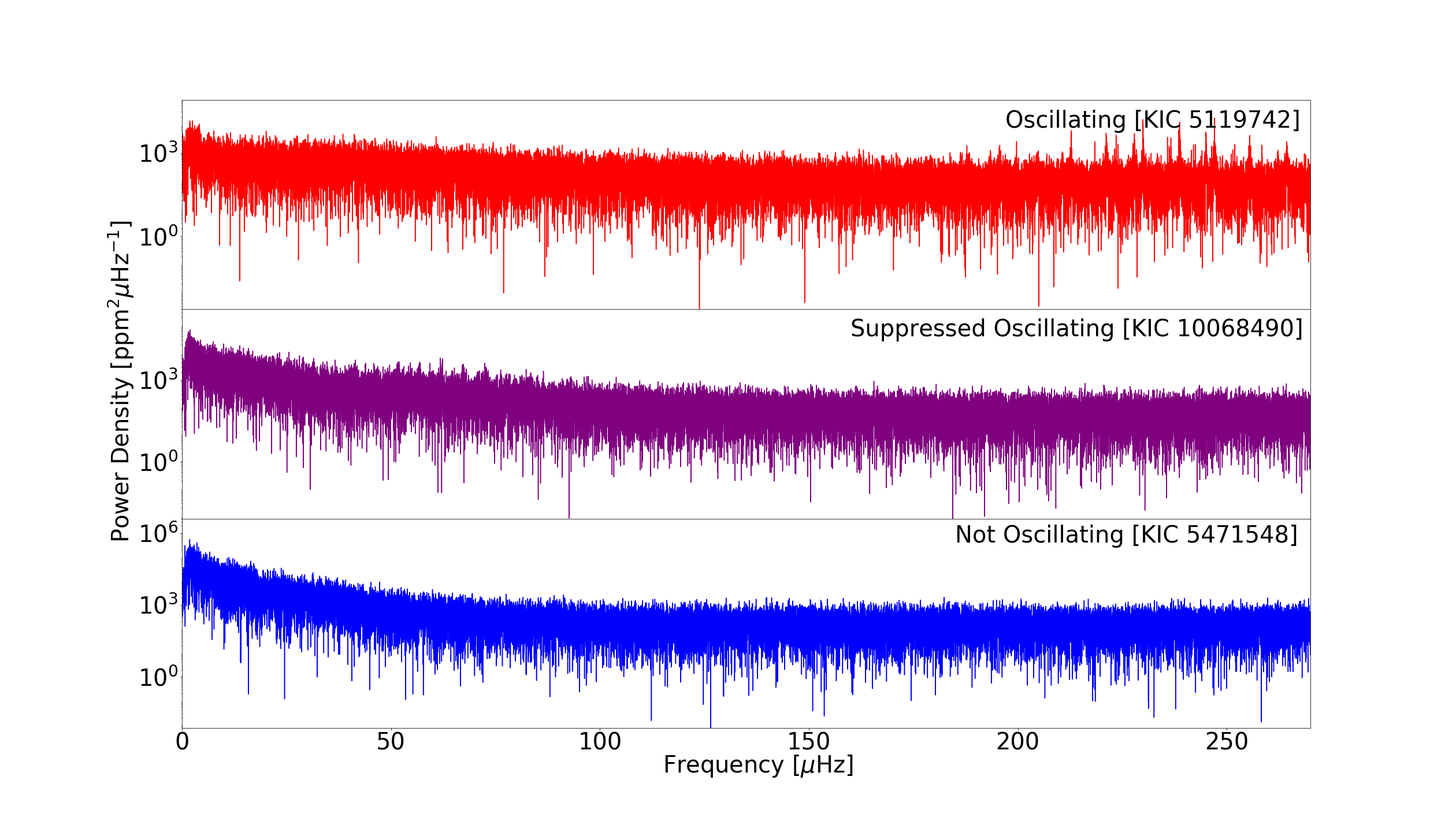}
\caption{Three \textit{Kepler} power spectra for an oscillating, suppressed and non-oscillating giant star. \textbf{Top:} KIC 5119742 with oscillations around 230$\mu$Hz. \textbf{Middle:} Suppressed oscillations in KIC 10068490; slight oscillations around 65$\mu$Hz. \textbf{Bottom:} No oscillations (KIC 5471548).}
\label{fig:giant_oscillations}
\end{figure*} 

\section{Target Selection}
\label{sec:target_selection}

Our sample contains \allstars{} dwarf (log(g) $>$3.5) and red giant (log(g) $<$3.5) stars \textit{predicted} to display solar-like oscillations. Oscillating red giants, as well as oscillating and anomalous\footnote{Throughout this work `anomalous' refers to stars with either suppressed oscillations or no oscillations.} dwarfs were randomly selected from the APOKASC catalog. Anomalous red giants were identified via visual inspection of spectroscopically confirmed red giants in the APOKASC catalog \citep{Pinsonneault2014}. Table \ref{tab:num_of_stars} organises the sample into sub-categories: dwarfs and giants; oscillating and anomalous.

We used stellar parameters to calculate the detection probability: the probability that oscillations would be detected above the \textit{Kepler} detection limit as described in \cite{Chaplin2011c}. Detection probability was calculated for all dwarfs using temperature values from the \textit{Kepler} Stellar Properties Catalog (KSPC) \citep{Mathur2017} and updated radii from Gaia Data Release 2 (DR2) \citep{GaiaCollaboration2018, Berger2018}. We then separated dwarfs into anomalous and oscillating groups, based on a limit of $\geq$90\% detection probability for oscillations. 

All giants have a detection probability of 100\%, based on their large pulsation amplitudes. There is a well understood relation between amplitude and frequency of maximum power ($\nu_{max}$) \citep{Huber2011}, so if a star has a much lower amplitude value than expected, we can define it as anomalous. To categorize giants we used amplitude and $\nu_{max}$ values from \cite{Yu2018}, which contains precise estimates of asteroseismic properties for 16,000 \textit{Kepler} red giants, some of which overlap with our sample. Figure \ref{fig:numax_amp} shows the data from \cite{Yu2018} and a fit to the $\nu_{max}$-amplitude relation with 3$\sigma$ limits. We defined all stars 3$\sigma$ \textit{below} the fit to be anomalous. Stars appearing 3$\sigma$ upward of the fit are likely high amplitude red giants, whose large amplitudes are thought to be due to triple systems, with a red giant and wide main-sequence binary contaminating the pixel aperture \citep{Colman2017}. 

For stars in the \cite{Yu2018} data set with $\nu_{max} > 200\mu Hz$, no amplitudes are listed. This is because at $\nu_{max} > 200\mu Hz$ it becomes difficult to fit the power spectrum background. These targets were marked as oscillating. Stars not included in the \cite{Yu2018} work were grouped based on a visual inspection of oscillations in their power spectra. Figure \ref{fig:giant_oscillations} illustrates example power spectra for three giant stars showing oscillations, suppression of oscillations, and no oscillations respectively. One star had no available power spectra and was not present in \cite{Yu2018} so it was marked as oscillating as is expected for giants.

\section{Observations and Data Reduction}
\subsection{Robo-AO}

We used the Robo-AO robotic laser AO system \citep{Baranec2014}, mounted on the 2.1m telescope at Kitt Peak, Arizona \citep{Jensen-Clem2018}, to obtain high angular resolution images of our full target sample (327 stars). Robo-AO observations took place between 2016 June 07 and 2017 May 28, across 20 nights, with 140 objects observed more than once to ensure high quality images. We used a total exposure time of 120s that enabled the detection of additional sources up to $\sim$6 magnitudes fainter than the target. We took all observations in the \textit{i}\textquotesingle{} filter (our stars range from magnitudes of 6.8 to 14 in i band). More information on the magnitude limits of observations at Kitt Peak can be found in \cite{Jensen-Clem2018}. 

We used the standard Robo-AO data reduction techniques described in \cite{Law2014}. Table \ref{short_long} lists all Robo-AO observations, including \textit{Kepler} Input Catalog (KIC) Identifier and i-band magnitudes. It also states whether a companion has been observed, either in this work or previously.

\subsection{NIRC2}

We used the NIRC2 infrared camera behind the Keck II AO system to confirm all the wide companion candidates, and obtain supplementary near-infrared photometry. We observed the targets on 2016 September 12, 13 and 2017 July 31. We operated NIRC2 in its 9.9 mas pixel$^{-1}$ mode which results in a field of view of $\sim$10\farcs0. We obtained 3-point dithered images for each star, with total exposure times ranging from 36s to 240s. We used the J, K' and PK-continuum filters (central wavelengths 1.248$\mu$m, 2.124$\mu$m, 2.2706$\mu$m and , respectively) choosing a filter consistent with achieving the best image of both star and companion. 

Each image from NIRC2 underwent sky subtraction and flat-field calibration. Flat-field frames were taken at the beginning of each night, and dark subtraction was performed with an unused quadrant of the detector. Each frame was corrected for bad pixels, and stacked to create a final image.

\subsection{ESPaDOnS}

ESPaDOnS is a high resolution echelle spectrograph at CFHT on Maunakea, Hawai'i. We chose a sub-sample of 34 stars, as observing constraints would not allow a multi-epoch survey of the entire target sample. The sub-sample contains both single stars (18) and stars with wide companions observed by Robo-AO in the imaging stage (15). We used ESPaDOnS to obtain at least three epochs of spectroscopy between 2017 and 2018, with the exception of KIC 893836, which was only observed twice but still included in analysis. Table \ref{tab:num_of_stars_2} organises this sample into the same categories as Table \ref{tab:num_of_stars}. Observations had an average signal-to-noise ratio (SNR) of $\sim$80 per frequency resolution element, at an average resolution of R$\sim$80,000. 

ESPaDOnS data is delivered to the user fully reduced using the Libre-ESpRIT reduction package \citep{Donati1997}. This package performs bias subtraction, flat-fielding, masking of bad pixels, wavelength calibration and spectrum extraction. The output provided contains several data analysis options: a continuum normalized spectrum, a corrected spectrum based on telluric lines or a combination of these options. We chose the continuum normalized data with the barycentric correction. Table \ref{Spectra} describes these observations as well as the results from the data analysis stage.

We included an eclipsing binary system from G14 in the spectroscopic sample: KIC 5308778. This star provided a test for whether our method was capable of revealing an RV scatter consistent with a spectroscopic binary. 
\section{Data Analysis}

\subsection{AO Companions}

\begin{table*}
\begin{longtable}{l || c c c c }
\caption{Full Robo-AO Observation List \label{tab:short_all_obs}}
\\
\hline
\textbf{KIC ID} & \textbf{Mag} & \textbf{Obs} & \textbf{Companion?\tablenotemark{1}} & \textbf{Category\tablenotemark{2}}\\[0.5ex]
\textbf{(KOI)} & \textbf{i'} & \textbf{Date} & &\\[0.5ex]
\hline
1430163&9.49&20160704 && DO \\[0.1ex]
1430239&10.39&20160704 && DA \\[0.1ex]
1571152&9.24&20160704 &Both& DA 
\\[0.1ex]
1576249&11.33&20160607 &Wide& DA 
\\[0.1ex]
1725815&10.71&20160607 &Both& DO 
\\[0.1ex]
1870433&12.34&20160704 && GA 
\\[0.1ex]
2140561&12.50&20160704 && GA 
\\[0.1ex]
2285032&11.25&20160704 && GA 
\\[0.1ex]
...&...&...&...&...
\label{short_long}

\end{longtable}
\tablenotetext{1}{Both = Wide and spectroscopic companion in the same system.}
\tablenotetext{2}{Short cadence targets: (DA) = anomalous dwarfs, (DO) = oscillating dwarfs. Long cadence targets: (GA) = anomalous giants, (GO) = oscillating giants. (G14) = \cite{Gaulme2014} red giant sample.}
\end{table*}



\begin{figure}[ht]
\centering
\includegraphics[width = 0.50\textwidth]{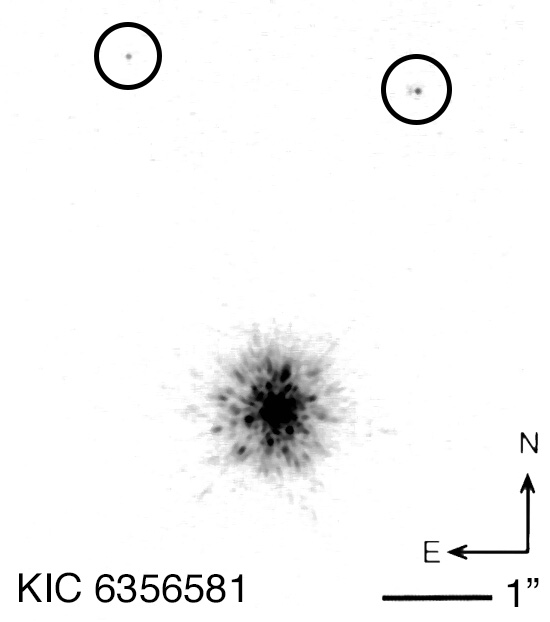}
\caption{Reduced NIRC2 image of triple system KIC 6356581. This image has been adjusted for contrast using SAO DS9, allowing the secondary and tertiary to be visible; both are circled. Neither of these companion stars is found to be physically associated i.e. they do not appear to be at a similar distance to the primary star.}
\label{fig:Discoveries_keck}
\end{figure}

\subsubsection{Companion Detection and System Confirmation}
\label{sec:ACDASC}

All detected wide candidate companion systems needed to be visually resolved in the full frame or PSF-subtracted image, in order to deduce system parameters using aperture photometry or PSF-fitting. To identify companions in the Robo-AO data we visually inspected the images for secondary stars with separations $\leq$4\farcs0, the size of a \textit{Kepler} pixel. Contaminating secondary stars may exist inside the [larger] \textit{Kepler} aperture, but these would be detectable in seeing-limited surveys and are therefore not included here. The search was aided by the Robo-AO data visualization and characterization GUI (Lamman et al. (in prep.)). 

\begin{table}[ht]
\centering
\caption{ESPaDOnS Sample Breakdown \label{tab:num_of_stars_2}}
\begin{tabular}{l|cc}
\hline
& \textbf{Dwarfs} & \textbf{Giants} \\
\hline
\textbf{Oscillating} & 18             & 5              \\
\textbf{Anomalous}   & 7              & 3              \\
\end{tabular}
\end{table}

We then confirmed our detections using an automated companion detection algorithm developed for the Robo-AO \textit{Kepler} Object of Interest (KOI) surveys (see \citealt{Ziegler2016}). A detection significance was found for each candidate companion by sampling and modeling the background noise level as a function of radial distance from the target star. We then slid an aperture of the diffraction-limited FWHM diameter along concentric annuli centered on the target star. Possible astrophysical detections are identified when the enclosed flux of the aperture becomes significantly greater than the local noise. In this sample of brighter stars, bright speckles can produce high-significance detections, which we discarded. We chose the significance value for which the companion pixel coordinates we manual identified matched with the pixel coordinates of the significance detection method.

\begin{figure*}[ht]
\centering
\includegraphics[width=0.85\textwidth]{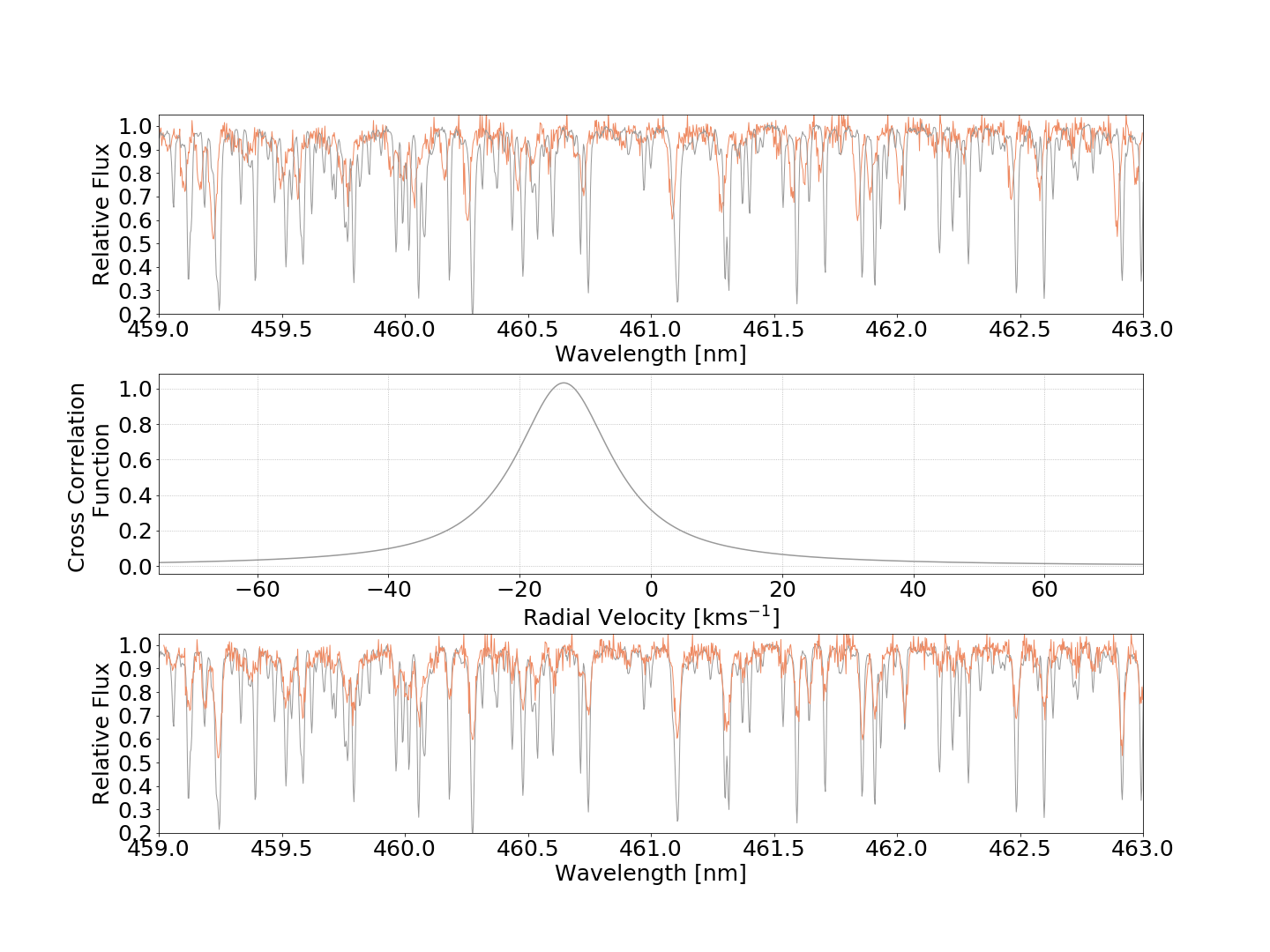}
\caption{ESPaDOnS spectra of a target, KIC 10124866, (orange) and standard star (gray). \textbf{Top:} Original spectra. \textbf{Middle:} Cross-correlation used to determine the radial velocity. \textbf{Bottom:} Shifted spectra once the cross-correlation has been applied.}
\label{fig:testing_rv_code}
\end{figure*}

All companions visually detected in the full frame images can be seen in Figure \ref{fig:appendix1} of the Appendix, whilst stars identified in the PSF-subtracted images can be seen in Figure \ref{fig:appendix2} of the Appendix. All detected companions from Robo-AO images were observationally confirmed using NIRC2. For three of these systems the NIRC2 observations revealed additional tertiary stars that remained undetected by Robo-AO, an example of which can be seen in Figure \ref{fig:Discoveries_keck}. 

We used these identified systems to calculate a companion fraction for anomalous and oscillating stars. A companion fraction is defined as the percentage of stars that have at least one discovered companion. This companion could be a wide companion (either physically associated or coincident) or a spectroscopic companion. When quoting companion fractions we used one of two uncertainties. Poisson errors apply only in the case of large samples so for N$>$100, errors are calculated in this way. For N$<$100 binomial errors were used. This latter method is taken from \cite{Burgasser2003} where statistical uncertainties are derived by constructing a probability distribution for the total sample size, N, and the number of binaries in the sample, n. The binomial distribution determines the probability of finding n binaries given the sample size and binary fraction.

\subsubsection{Separation and Position Angle}
We calculated separation and position angle between the primary and secondary stars with the same technique as \cite{schonhutstasik2017}, adapted for Robo-AO at Kitt Peak. We tested this method with Palomar data from \cite{baranec2016} and Kitt Peak data from \cite{Ziegler2018} reproducing separation, position angle and uncertainty values. For pairs too close to resolve in the reduced image, coordinates from a PSF-subtracted image were used by shifting them to the frame of the reduced image.

Pixel coordinates were determined using the Aperture Photometry Tool\footnote{http://www.aperturephotometry.org} (APT) \citep{Laher2012}, except for KIC 3430893 where APT could not lock on correctly to either the full frame or PSF-subtracted image. In this case, coordinates were determined using SAO DS9\footnote{http://ds9.si.edu/site/Home.html}, by taking an average of manual measurements of the central pixel for the star in question. We also manually determined the separation for tertiary companions identified in NIRC2, with the error equivalent to the size of a NIRC2 pixel, i.e. 0\farcs01. 

\subsubsection{Contrast Ratios and Amplitude Dilution}

Flux ratios were calculated using PSF-photometry, designed using the Astropy\footnote{http://www.astropy.org} module Photutils\footnote{https://photutils.readthedocs.io/en/stable/}. We calculated the ratio by using Gaussian models to fit the centroid coordinates of each star, determining their relative flux ratio.

For magnitude differences, as well as individual magnitudes and fluxes, we used the method from \cite{schonhutstasik2017} for both Robo-AO and NIRC2 data. To find individual magnitudes we compared the flux ratio to the total magnitude of the system, taken from the KIC. We used i-band for the total system magnitudes of the Robo-AO images and different bands for the NIRC2 images depending on the filter, (K for K' and PK-continuum, J for J). We note that i-band corrections may be overestimated for widely separated sources, given the typical $\sim$2\farcs5 resolution of KIC photometry \citep{Brown2011}.

Amplitude dilution is defined as the percentage of flux observed from the system that is a result of the secondary star:

\begin{equation}
    A = \frac{F_2}{F_1 + F_2} \times 100
\end{equation}
with $F_1$ and $F_2$ corresponding to primary and secondary fluxes, respectively. The effect of amplitude dilution is larger in triple systems containing two wide companions. When only one companion accompanies the primary star, the companion can only dilute the flux by a maximum of 50\%. However, in a three star system, with two extra stars, the maximum amplitude dilution is 67\%, with each star contributing a third of the flux.

\begin{figure}
\centering
\includegraphics[width = 0.5\textwidth]{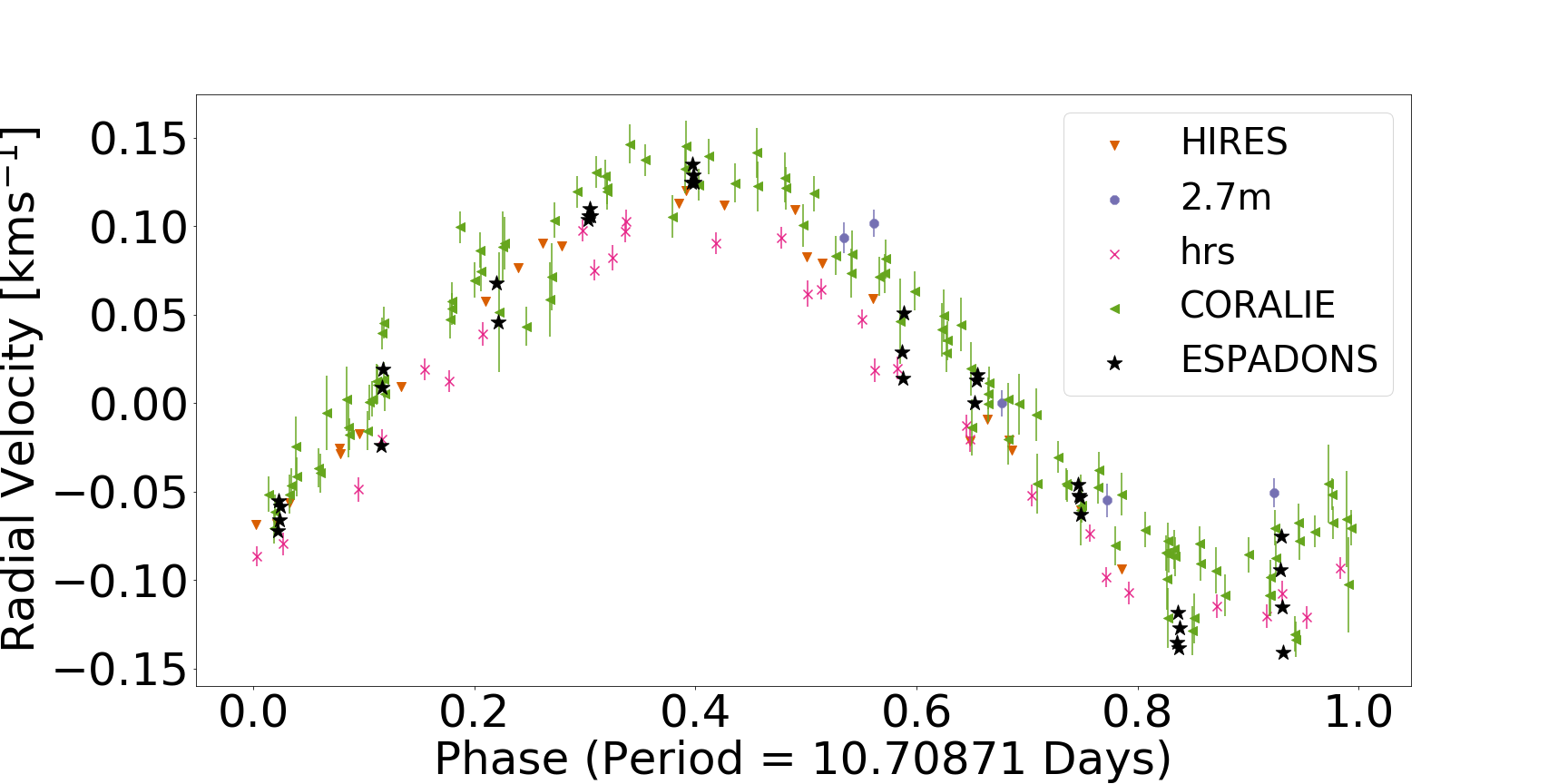}
\caption{Reproduction of the planet phase curve from \cite{Hinkel2015} demonstrating that HIP72339 hosts a hot Jupiter. Each marker and color represents a different set of observations from a different telescope. Radial velocities determined in this work (using data from \cite{Fares2013}) are marked by black stars.}
\label{fig:planet_curve}
\end{figure}

\begin{figure}
\centering
\includegraphics[width = 0.5\textwidth]{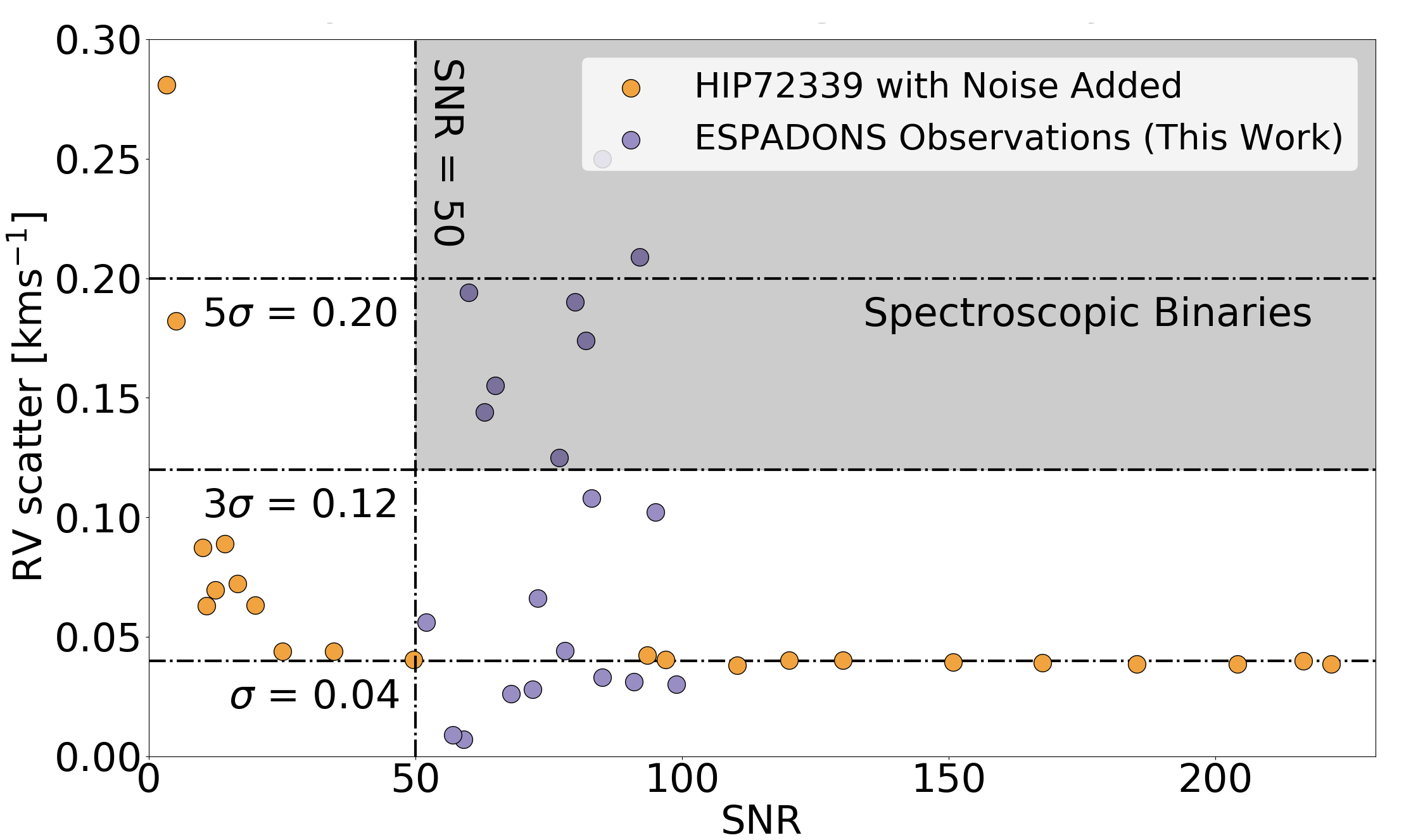}
\caption{Radial velocity scatter versus SNR. Orange symbols represent the standard star with a varying SNR, simulated by adding Gaussian noise to the spectra. Purple symbols illustrate our target sample. The horizontal dashed line marks the 1$\sigma$ limit. Above 3$\sigma$ we define our targets to be RV variable due to a stellar companion (shaded area). The y-axis has been truncated to not include the larger values of RV scatter.}
\label{fig:interpolate}
\end{figure}

\subsubsection{Companion Characterization}

Primary spectral types were taken from the SIMBAD database\footnote{http://simbad.u-strasbg.fr} \citep{Wenger200}, or if unavailable, were inferred from the effective temperature of the primary star using Table 5 in \cite{Kraus2007}. This assumes that the companion had a negligible effect. Parameters for the wide systems can be found in Table \ref{CompDet}. Statistical analysis of those systems is presented in Section \ref{sec:results}.

Where available, we used Gaia DR2 for updated radii \citep{Berger2018} and distance measurements for both stars. We used primary target RA and Dec from SIMBAD or KSPC and a 5\farcs0 circluar aperture, to search for detected companions in the DR2 database. Separation was calculated using RA and Dec values for any other stars located in the aperture, to confirm whether the companion identified with Robo-AO was also identified with Gaia. Nine primary targets had no Gaia radius so KSPC values were used. 

\begin{figure*}
\centering
\includegraphics[width=0.7\textwidth]{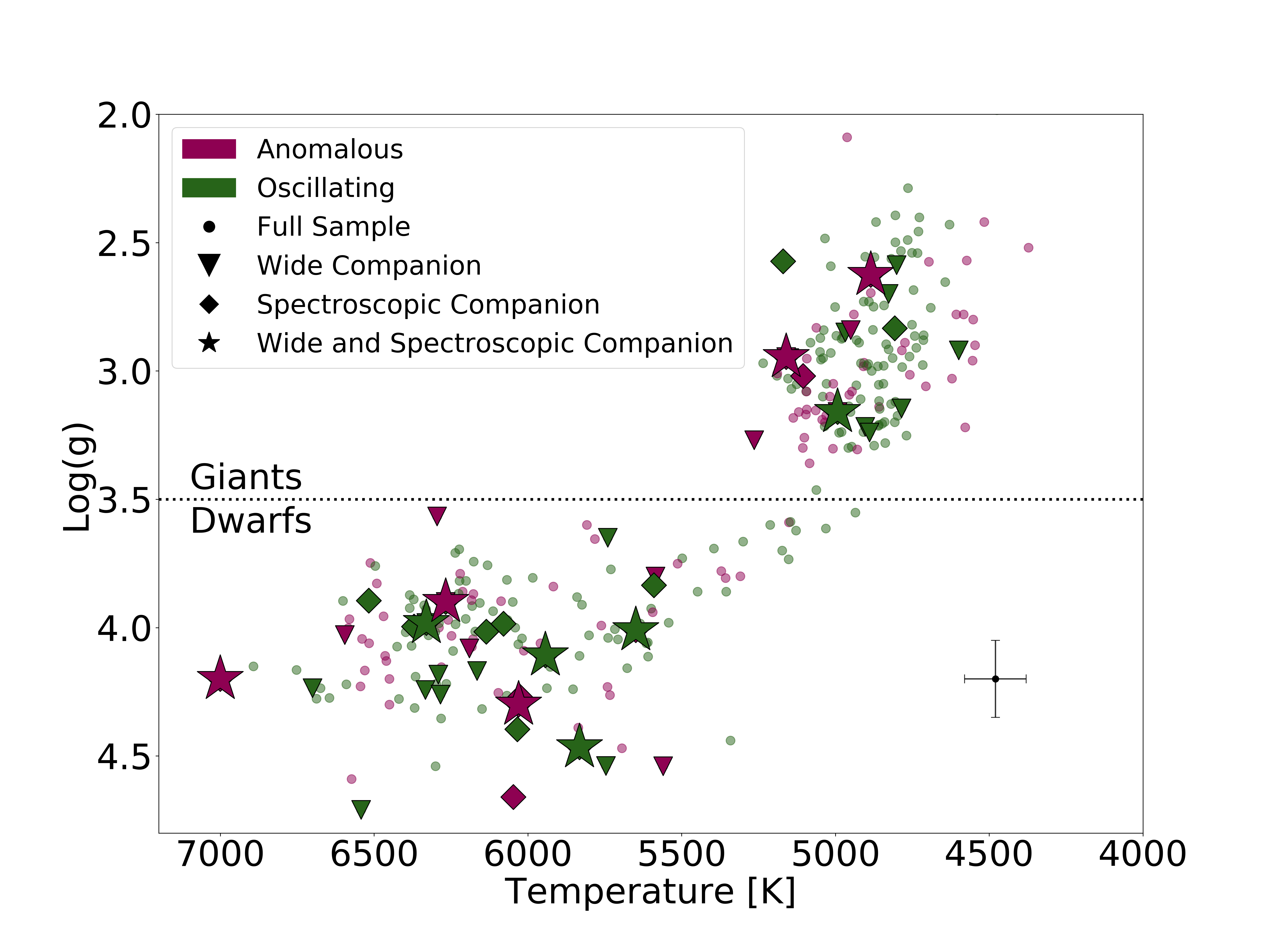}
\caption{H-R diagram showing the full sample of oscillating (green) and anomalous (purple) stars with surface gravity on the y-axis and temperature on the x-axis (both from KSPC). The error shown (100K for temperature and 0.15 dex for surface gravity) is a typical uncertainty for the whole sample. A line separates the dwarfs and giants at log(g) = 3.5. The circles in this plot represent the targets for which no companions were found. Targets with a wide companion are shown by a downward facing triangle, the spectroscopic binaries are marked by a diamond and targets with both a wide and spectroscopic companion are marked by a star.}
\label{fig:final_hr}
\end{figure*}

\subsubsection{Physical Association}
Physical association between the two stars in a system was determined by calculating whether their distances agree within their uncertainties to 1$\sigma$. If the distance to a star was not available in Gaia, but we had a K' or PK-continuum band NIRC2 observation, we used the method described in \cite{Atkinson2017} (hereafter A17). This method uses broadband photometry to determine radii, spectral types and distances to stars. If possible, we used both A17 and Gaia to compare computed distances and determine the accuracy of the A17 model. If the distances given by A17 and Gaia agreed within uncertainty, we adopted the spectral type and radii for the secondary given by A17. The A17 model does not discriminate between dwarfs and giants and in these cases will give an incorrect distance, therefore these radii and spectral type were not included. These values can be found in Table \ref{CompDet3}. 

Both \cite{Ziegler2018b} and \cite{Hirsch2017} found that most binary systems with separations of $\leq$1\farcs0 are physically associated. We adopt this for all our binaries at $\leq$1\farcs0 without derived distances (unavailable in Gaia and no appropriate NIRC2 images).

\subsection{Spectroscopic Companions}
\subsubsection{Radial Velocities}

We used spectroscopy to identify binaries too close to be resolved using AO imaging. We detected companion systems by measuring the scatter in radial velocity (RV) of a star over at least three epochs.

To determine the RV of each observation we used a cross-correlation function (as implemented in pyasl\footnote{https://github.com/sczesla/PyAstronomy} with a step size of 0.001kms$^{-1}$) to compare to a standard star, as demonstrated in Figure \ref{fig:testing_rv_code}. We found the absolute RV values by subtracting the RV of the standard (in this case -11.85kms$^{-1}$). The RV value for an observation was found from the peak of the cross-correlation curve. We took the RV scatter as the standard deviation of all the RV values for a star. 

For wavelength ranges, we chose 459nm - 463nm for cooler stars and 490nm - 495nm for hotter stars, containing as many lines as possible without including strong telluric lines. We were careful to include a similar number of lines in both hot and cold stellar spectra, so we could assume a standard RV uncertainty amongst all stars. 

\subsubsection{The Standard Star}
To choose a standard, we tested stars from \cite{Soubiran2013} who created a catalog of 1420 RV standards to calibrate spectroscopic measurements for Gaia. Our standard star needed multiple ESPaDOnS observations (taken from the Canada Astronomy Data Centre\footnote{http://www.cadc-ccda.hia-iha.nrc-cnrc.gc.ca/en/}) and a high SNR. 

The best candidate to meet these criteria was HIP72339, which incidentally hosts a hot Jupiter first discovered in \cite{Udry2000}. We used the presence of a planet to test the accuracy of our cross-correlation function, by reproducing the phase curve of the planet from \cite{Hinkel2015} (see Figure \ref{fig:planet_curve}). Our ability to identify this hot Jupiter is evidence that our method is adequate for discovering spectroscopic binaries. 

As HIP72339 is not in the \textit{Kepler} field, we also ran a number of cross-correlations using \textit{Kepler} stars from our sample, in order to ensure that error would not be introduced using this standard.

\subsubsection{RV Uncertainties}
To determine which values of RV scatter should be considered indicative of a spectroscopic binary, we calculated a lower limit for the RV scatter, by building a relationship between RV scatter and SNR. 

The average SNR for the HIP72339 observations was higher than our targets, allowing us to use the standard to dictate a lower limit. We plotted the SNR versus RV scatter (with the planet subtracted out) for observations of HIP72339 with various quality levels. We varied the quality level by adding random noise from a Gaussian distribution. We obtained the SNR from the continuum of the spectra, by measuring the mean of the points around 606nm, where there are no spectral lines. We obtained the standard scatter by cross-correlating against another observation of the standard and plotted the RV scatter as a function of SNR for each iteration of the observation + noise. The result of this can be seen in Figure \ref{fig:interpolate}.

We then calculated a SNR for each set of target observations and overplotted these on to the values of HIP72339.  The values of SNR versus scatter for HIP72339 stayed consistent down to $\sim$30. As all our target observations had an SNR$>$50, we chose $\sigma$=0.04kms$^{-1}$ to be our lower limit. This corresponds to the mean of the HIP72339 observations with SNR$>$30. This value gave us a 3$\sigma$ detection of 0.12kms$^{-1}$ and a 5$\sigma$ detection of 0.20kms$^{-1}$.

\section{Results}
\label{sec:results}
\subsection{All Companions}

All the systems identified in this survey are summarized in Figure \ref{fig:final_hr}, which shows no trend in where the stars lie on the H-R diagram. 

\begin{figure}[ht!]
\centering
\includegraphics[width = 0.5\textwidth]{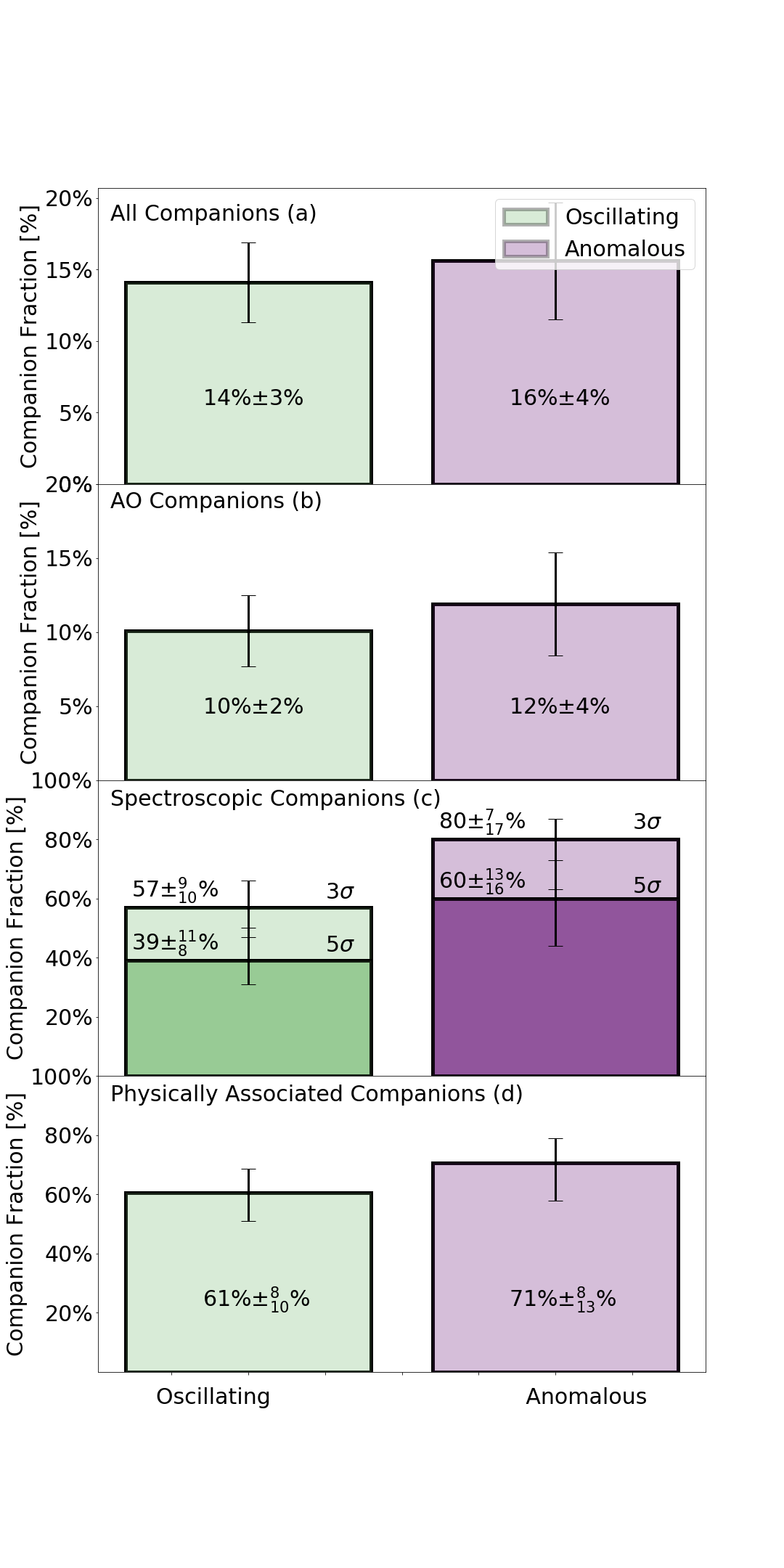}
\caption{Companion fractions for oscillating (green) and anomalous (purple) systems. \textbf{Top:} All companion systems. \textbf{Second from Top:} AO/wide companions. \textbf{Third from Top:} Spectroscopic companions and \textbf{Bottom:} Physically associated companions. Uncertainties are calculated using binomial or Poisson statistics depending on the sample size.}
\label{fig:multi_osc_vs_not_2}
\end{figure}

We compared the overall companion fraction of oscillating and anomalous stars (Figure \ref{fig:multi_osc_vs_not_2}(a)) and found that they agree within uncertainty, suggesting no difference in multiplicity between the groups. This companion fraction includes any system with \textit{at least} one wide or spectroscopic companion; systems with multiple companions are only considered once.

\begin{table*}
\centering
\begin{longtable}{l || c c c c c c c }
\caption{Radial Velocity Shifts for Spectroscopic Data \label{Spectra}}\\

\hline

\textbf{KIC ID} & \textbf{Julian} & \textbf{Radial} & \textbf{Sigma} & \textbf{Standard} & \textbf{Signal to} \\
& \textbf{Date} & \textbf{Velocity} & \textbf{Likelihood} & \textbf{Deviation} & \textbf{Noise Ratio} \\
& \textbf{[Modified]}&\textbf{ [kms$^{-1}$]}&&\textbf{ [kms$^{-1}$]}& \\
\hline

1571152 & 57884.468353 & 7.150 & 5 & 2.131 & 93 [86] \\
& 58005.2196723 & 9.437 &&& 92 \\
& 58297.6207027 & 7.149 &&& 113 \\
& 58360.4760237 & 9.187 &&& 61 \\
& 58391.3431338 & 3.487 &&& 88 \\
1725815  & 57884.5152016 & 24.778 & 5 & 2.433 & 84 [87] \\
& 57972.385056 & 23.241 &&& 81 \\
& 58321.3778602 & 19.024 &&& 96 \\
3115178  & 57879.564732 & -26.694 & 1 & 0.056 & 60 [52] \\
& 57979.3868736 & -26.674 &&& 42 \\
& 58298.580257 & -26.567 &&& 61 \\
... & ... & ... & ... & ... & ...

\end{longtable}
\end{table*}

\begin{figure*}[ht!]
\centering
\includegraphics[width=\textwidth]{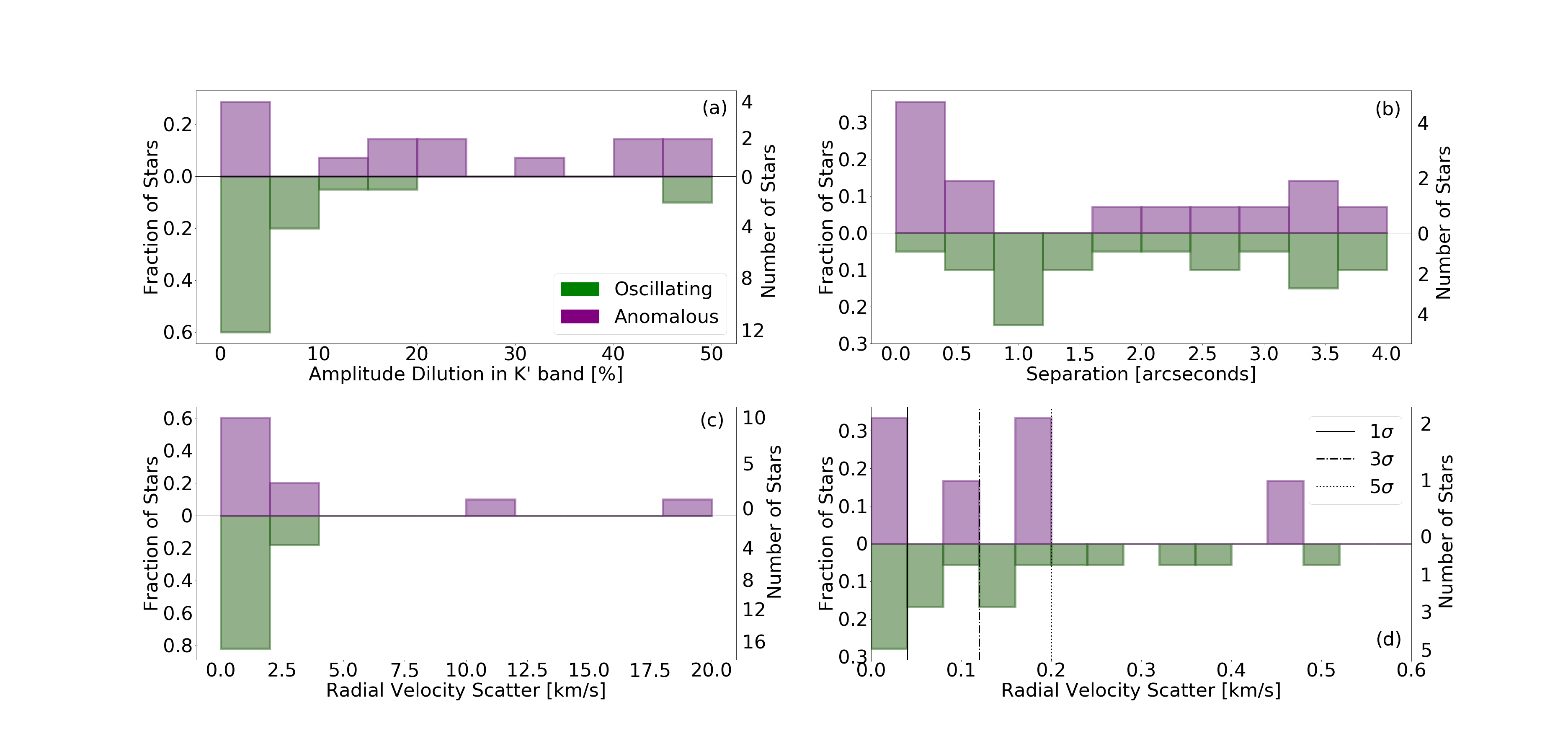}
\caption{Plots showing the distribution of companion systems in the oscillating and anomalous groups as a function of system parameters. \textbf{Top Left:} Distribution as a function of amplitude dilution in the K'-band (\%). \textbf{Top Right:} Distribution as a function of separation from the primary star (arcseconds). \textbf{Bottom Left:} Distribution as a function of the radial velocity scatter (kms$^{-1}$). \textbf{Bottom Right:} Distribution as a function of radial velocity scatter up to 0.6kms$^{-1}$ (kms$^{-1}$). Also showing 1, 3 and 5$\sigma$ lines.}
\label{fig:multi_osc_vs_not}
\end{figure*}

\subsection{AO/Wide Companions}

We identified 34 systems with at least one wide companion (11\%$\pm$2\% companion fraction) and 18 systems containing a companion which may be physically associated (6\%$\pm$1\% companion fraction). We find four systems with more than one wide companion, however none with more than one physically associated wide companion. These statistics do not include the G14 sample.

The system with smallest separation is KIC 2568519 at 0\farcs16 $\pm$ 0\farcs08, almost at the diffraction limit of Robo-AO. Three triple systems were observed: KIC 3221671, KIC 6356581 and KIC 8983847. KIC 5123145 is a quadruple system. 

\subsection{Spectroscopic Companions}
We found 15 spectroscopic binaries, giving a companion fraction of 41$\%^{+11\%}_{-9\%}$ when considering a 5$\sigma$ lower limit. The highest RV scatter was 19.745kms$^{-1}$ for KIC 5308778, a known G14 binary. The second largest was KIC 11551430 with an RV scatter of 19.0kms$^{-1}$. Five systems were found to have both a wide and spectroscopic companion (to 5$\sigma$) and four where both components are physically associated. 
\subsection{Comparison to Other Surveys}

The companion fraction for the oscillating dwarf group (18\%$\pm$5\%) is lower than the value from \cite{Raghavan2010}, who find a companion fraction of $\sim$45\% for FGK dwarfs. Our lower fraction is likely due to the fact our survey truncates possible binaries at 4\farcs, whereas some methods in \cite{Raghavan2010} identify binaries out to 200\farcs. We were unable to compare our anomalous systems to \cite{Raghavan2010} as they were chosen for possible binarity and therefore present a selection bias.

To compare our companion fraction with a similar sample also observed with Robo-AO, we used values from their KOI survey \citep{Ziegler2018}. We combined the oscillating dwarfs in our work with a sample of 99 oscillating \textit{Kepler} dwarfs and subgiants from \cite{schonhutstasik2017} to produce a companion fraction representing dwarf and subgiant solar-like oscillators. Excluding giants makes the asteroseismic sample more consistent with the KOI sample. We found the companion fraction of the KOI survey (14.5\% $\pm$ 0.5\%) to be in agreement with ours (13\% $\pm$ 3\%).

We were unable to compare the dwarf and giant samples to one another due to the bias in completeness for the oscillations and for the binary detection. For the remainder of this analysis the oscillating and anomalous groups contain both dwarfs and giants. This should not effect overall binary fraction as there are roughly the same number of dwarfs and giants in each category.

\section{Discussion}
\subsection{Wide Companions}

Figure \ref{fig:multi_osc_vs_not_2}(b) compares the companion fraction of wide binaries in the oscillating and anomalous groups, showing that anomalous stars are no more likely to have a wide companion than oscillating stars. This calculation considers each system only once, regardless of the number of companion stars within it.

Figure \ref{fig:multi_osc_vs_not}(a) shows the distribution of wide companions as a function of their amplitude dilution in K'-band. For amplitude dilutions $>$10\%, more systems belong to the anomalous group. This suggests there may be a lower limit to the amount of amplitude dilution necessary, in order to reduce observed oscillations below the \textit{Kepler} detection limit. 

\begin{figure}[ht!]
\centering
\includegraphics[width = 0.45\textwidth]{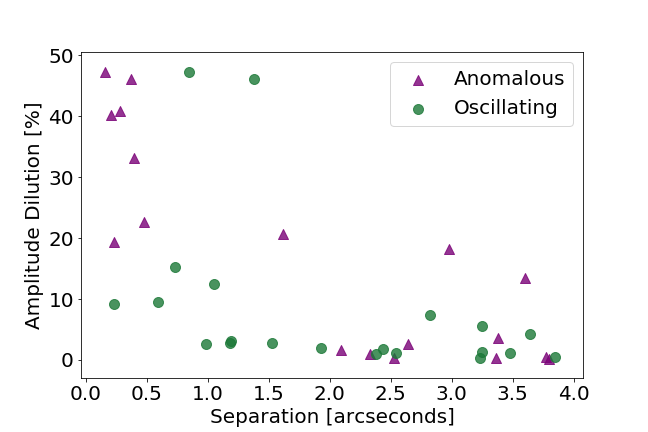}
\caption{Amplitude dilution (in K' band) as a function of separation (arcseconds) for both the oscillating and anomalous wide companion systems. Purple triangles represent anomalous wide companion systems and green circles represent oscillating wide companion system.}
\label{fig:amp_sep}
\end{figure}

Figure \ref{fig:multi_osc_vs_not}(b) shows the distribution of wide companions as a function of their separation from the primary star. It shows more companions to anomalous stars at close separations of $\leq$1\farcs. Closer companions are more likely to be physically associated equal mass companions \citep{Raghavan2010} than companions at a wider separation. This suggests a higher value of amplitude dilution for closely separated systems, as the secondary star will be contributing a greater flux. In Figure \ref{fig:amp_sep} we have shown that at $>$2\farcs0 separation, generally all amplitude dilutions are $<$10\%. This reiterates that larger values of amplitude dilution are more likely at close separations.

\subsection{Spectroscopic Companions}

Figure \ref{fig:multi_osc_vs_not_2}(c) compares the companion fractions for the spectroscopic sample between the oscillating and anomalous groups. The companion fractions agree for both the 3$\sigma$ and 5$\sigma$ limit.

Figure \ref{fig:multi_osc_vs_not}(c) and (d) demonstrate the distribution in radial velocity scatter for the spectroscopic companions found in both the oscillating and anomalous samples. We tentatively observe higher RV scatter in the anomalous stars. 

As this work contains a large number of both oscillating and anomalous stars, we can assume that the companion mass and inclination distribution are the same for both groups. Therefore, we can interpret this finding as demonstrating a higher fraction of close companions in anomalous stars. 

\subsection{Triple Systems With Both a Wide and Spectroscopic Companion}

We observed both the G14 sample with direct imaging and a sub-sample of the wide binaries with spectroscopy. This allowed us to further investigate the results of \cite{Tokovinin2006}, who found that the presence of a spectroscopic binary is indicative of a wide companion. 

We did not find any wide companions to the G14 stars. This could be explained if our sensitivity did not achieve the necessary contrast ratio. Alternatively, \cite{Tokovinin2006} states that the probability of a wide tertiary companion drops from 96\% to 34\% with an orbital period increase from 3 to 12 days. The minimum period in the G14 sample is $\sim$15 days, therefore the lack of wide companions could indicate the probability of a wide tertiary companion continues to decrease with increased orbital period. 

We found that $\sim$50\% of the 3$\sigma$ spectroscopic binaries have a \textit{physically associated} tertiary (as opposed to a possible coincident wide companion or no tertiary companion at all). This rate is in agreement with the value for stars without a spectroscopic binary but with a physically associated wide companion. Therefore, it is no more likely that we would find spectroscopic binaries in wide systems.

\section{Conclusions}
\label{sec:conclusion}

We observed \allstars{} asteroseismic \textit{Kepler} stars with AO imaging and 34 with spectroscopy, to investigate whether stellar multiplicity is related to the suppression of solar-like oscillations. Our main conclusions are as follows: 

\begin{itemize}
\item We do not see a significant difference in companion fraction for wide companions between oscillating and anomalous stars (10\% $\pm$ 2\% and 12\% $\pm$ 4\% respectively; see Figure \ref{fig:multi_osc_vs_not_2}(b)). However, companions at separations of $<$0\farcs5, and demonstrating an amplitude dilution $>$10\% are all anomalous. This suggests that the presence of a wide companion is not enough to assume a star will have suppressed oscillations and it could indicate a threshold below which the presence of excess flux will not reduce the amplitude enough to create a non-detection, although it may still contribute along with other factors (i.e. triple systems).

\item We find tentative evidence for a higher fraction of spectroscopic binaries between anomalous and oscillating stars ($60\%\pm3\%$ and $39\%^{+11\%}_{-8\%}$ respectively (5$\sigma$); see Figure \ref{fig:multi_osc_vs_not}(c)), as inferred from the radial velocity scatter measured over multiple epochs. This would be in line with the suggestion by \cite{Gaulme2014} that tidal interactions in close eclipsing binary systems may suppress the convective driving of solar-like oscillations. Further observations of a larger number of systems with an extended baseline of spectroscopic follow-up will be required to confirm this result.

\item Although companion systems are a likely mechanism for the non-detection on oscillations in some stars, it is probably not the only mechanism. There are still 75 anomalous stars in this sample (109 total) for which no companion has been detected. This does not rule out multiplicity as it is likely that not all companions were discovered in this work. This could be due to separations too close for AO imaging to resolve but too far out for spectroscopy. It could also mean their oscillation suppression comes from another source, such as increased levels of stellar activity. A study of stellar activity in our sample is beyond the scope of this paper.

\item For all physically associated wide companion systems that were also surveyed with ESPaDOnS, $\sim$50\% contain a spectroscopic binary to 3$\sigma$, consistent with the spectroscopic companion fraction for single systems. We did not find any wide companions to the G14 sample. This may be because the probability of a wide tertiary companion decreases with the increased orbital period of the close binary \cite{Tokovinin2006}. In the G14 sample all the eclipsing binaries have orbital periods $\geq$15 days, larger than those in the \cite{Tokovinin2006} sample.
\end{itemize}

The Transiting Exoplanet Survey Satellite (TESS) mission is observing an order of magnitude more asteroseismic stars than \textit{Kepler}. TESS has 21\farcs{} pixels, so determining multiplicity will be even more crucial as more blended binaries in the aperture can add flux to the primary light curve. Robo-AO will be used to vet the majority of TESS candidate exoplanet host stars \citep{ZieglerGaia}, and in a similar process, can also be used to find candidate stellar companions to asteroseismic stars. Thorough and timely follow-up will be required to reassess amplitude dilution in this much larger sample of asteroseismic targets.

\acknowledgments

D.H. acknowledges support by the National Science Foundation (AST-1717000) and the National Aeronautics and Space Administration under Grant NNX14AB92G issued through the Kepler Participating Scientist Program. (80NSSC19K0597).

C.B. acknowledges support from the Alfred P. Sloan Foundation.

The Robo-AO instrument was developed with support from the National Science Foundation under grants AST-0906060, AST-0960343, and AST-1207891, IUCAA, the Mt. Cuba Astronomical Foundation, and by a gift from Samuel Oschin.

The Robo-AO team thanks NSF and NOAO for making the Kitt Peak 2.1-m telescope available. We thank the observatory staff at Kitt Peak for their efforts to assist Robo-AO KP operations. The authors are honored to be permitted to conduct astronomical research on Iolkam Du'ag (Kitt Peak), a mountain with particular significance to the Tohono O'odham Nation. Robo-AO KP is a partnership between the California Institute of Technology, the University of Hawai‘i, the University of North Carolina at Chapel Hill, the Inter-University Centre for Astronomy and Astrophysics (IUCAA) at Pune, India, and the National Central University, Taiwan. The Murty family feels very happy to have added a small value to this important project. Robo-AO KP is also supported by grants from the John Templeton Foundation and the Mt. Cuba Astronomical Foundation. 

Some data are based on observations at Kitt Peak National Observatory, National Optical Astronomy
Observatory (NOAO Prop. ID: 15B-3001), which is operated by the Association of Universities for Research in Astronomy (AURA) under cooperative agreement with the National Science Foundation.

Some of the data presented herein were obtained at the W. M. Keck Observatory, which is operated as a scientific partnership among the California Institute of Technology, the University of California and the National Aeronautics and Space Administration. The Observatory was made possible by the generous financial support of the W. M. Keck Foundation.

Based on observations obtained at the Canada-France-Hawai`i Telescope (CFHT) which is operated by the National Research Council of Canada, the Institut National des Sciences de l'Univers of the Centre National de la Recherche Scientifique of France, and the University of Hawai`i.

The authors wish to recognize and acknowledge the very significant cultural role and reverence that the summit of Maunakea has always had within the indigenous Hawai`ian community.  We are most fortunate to have the opportunity to conduct observations from this mountain.

This research used the facilities of the Canadian Astronomy Data Centre operated by the National Research Council of Canada with the support of the Canadian Space Agency.

This research has made use of the SIMBAD database, operated at CDS, Strasbourg, France.

This research has made use of NASA's Astrophysics Data System Bibliographic Services.

We thank Dani Atkinson, Claire Moutou, Pascal Fouqué and Nadine Manset for vital communications and important advice. We acknowledge the Exoplanet Follow-up Observing Program (ExoFOP) and \textit{Kepler} Asteroseismic Science Operations Center (KASOC) databases, Python modules cv2 and pillow. 

\facilities{KPNO:2.1m (Robo-AO), Keck:II (NIRC2-NGS), CFHT (ESPaDOnS)}

\software{Astropy \citep{Astropy2013}, Pyasl (\url{https://pyastronomy.readthedocs.io/en/latest/}), Libre-Esprit \citep{Donati1997}, Aperture Photometry Tool (APT) \citep{Laher2012}, DS9 \citep{Joye2003,SAO2000}}


\bibliographystyle{compact_bib.bst}
\bibliography{References}


\clearpage
    \centering
\begin{longtable}{l || c c c c c c c c c}
\caption{Detected Companion Systems \label{CompDet}}\\

\hline\\

\textbf{KIC ID} & \textbf{Separation \tablenotemark{1}} & \textbf{Position} & \textbf{Magnitude} & \textbf{Magnitude} & \textbf{i\textquotesingle{} Detection} & \textbf{System} & \textbf{Amplitude} & \textbf{Amplitude}\\
& \textbf{(\arcsec)} & \textbf{Angle} & \textbf{Difference} & \textbf{Difference} & \textbf{Significance} & \textbf{Spectral} & \textbf{Dilution} & \textbf{Dilution}\\
&  & \textbf{($^{\circ}$)} & \textbf{i\textquotesingle{}} & \textbf{K'} & \textbf{ ($\sigma$)} & \textbf{Type} & \textbf{i\textquotesingle{} (\%)} & \textbf{K' (\%)} \\

\\
\textbf{1571152} & 0.40 $\pm$ 0.06 & 126 $\pm$ 2 & 0.91 $\pm$ 0.11 & 0.76 $\pm$ 0.18 & 7.99 & F2V & 30.20 $\pm$ 0.40 & 33.24 $\pm$ 9.16 \\
\textbf{1576249} & 0.28 $\pm$ 0.06\tablenotemark{1} & 164 $\pm$ 7 & 0.62 $\pm$ 0.21 & 0.40 $\pm$ 1.90 & 5.66 & F7V & 36.20 $\pm$ 0.45 & 40.94 $\pm$ 136.44 \\
\textbf{1725815} & 3.64 $\pm$  0.06  &  81 $\pm$ 2 & 3.81 $\pm$ 0.20 & 3.22 $\pm$ 0.17\tablenotemark{3} & 8.67 & F7V & 2.91 $\pm$ 0.05 & 4.19 $\pm$ 1.31\\
\textbf{2568519} & 0.16 $\pm$ 0.08\tablenotemark{1} & 74 $\pm$ 12 & 0.94 $\pm$ 0.22 & 0.29 $\pm$ 0.20 & 8.42 & F7V & 27.70 $\pm$ 3.99 & 47.35 $\pm$ 15.57 \\
\textbf{3123191} & 0.73 $\pm$ 0.06\tablenotemark{1} & 122 $\pm$ 3 & 1.91 $\pm$ 0.21 & 1.86 $\pm$ 0.16 & 5.24 & F7V & 14.47 $\pm$ 0.24 & 15.26 $\pm$ 3.78 \\
\textbf{3221671(1)} & 1.62 $\pm$ 0.06 & 217 $\pm$ 2 & 2.56 $\pm$ 0.20 & 1.46 $\pm$ 0.17 & 7.73 & F5V & 8.66 $\pm$ 0.15 & 20.63 $\pm$ 5.3 \\
\textbf{3221671(2)} & 2.09 $\pm$ 0.01 & ... & ... & 4.43 $\pm$ 0.17 & ... & ... & ... & 1.66 $\pm$ 0.42 \\
\textbf{3430893} & 1.18 $\pm$ 0.06 & 215 $\pm$ 2 & 3.37 $\pm$ 0.21 & 3.88 $\pm$ 0.16\tablenotemark{3} & $<$3 & F7V & 4.31 $\pm$ 0.08 & 2.73 $\pm$ 0.69 \\
\textbf{3643774} & 2.38 $\pm$ 0.06 & 106 $\pm$ 2 & 5.05 $\pm$ 0.19 & 5.04 $\pm$ 0.17 & 7.98 & G1V & 0.87 $\pm$ 0.02 & 0.96 $\pm$ 0.25 \\
\textbf{4260884} & 0.48 $\pm$ 0.06 & 177 $\pm$ 4 & 1.06 $\pm$ 0.21 & 1.33 $\pm$ 0.17 & 5.14 & K3III & 27.30 $\pm$ 0.38 & 22.72 $\pm$ 5.89 \\
\textbf{4914234} & 3.85 $\pm$ 0.06 & 165.8 $\pm$ 1.6 & 6.27 $\pm$ 0.22 & 5.81 $\pm$ 0.16 & 7.21 & K3III & 0.31 $\pm$ 0.01 & 0.47 $\pm$ 0.11 \\
\textbf{4999260} & 0.99 $\pm$ 0.06 & 342 $\pm$ 2 & 2.33 $\pm$ 0.21 & 3.96 $\pm$ 0.16\tablenotemark{3} & 4.09 & K3III & 10.50 $\pm$ 0.18 & 2.55 $\pm$ 0.63 \\
\textbf{5123145(1)} & 2.64 $\pm$ 0.06 & 230 $\pm$ 2 & 4.03 $\pm$ 0.20 & 3.91 $\pm$ 0.17 & 5.52 & K3III & 2.38 $\pm$ 0.04 & 2.65 $\pm$ 0.70 \\
\textbf{5123145(2)} & 0.23\tablenotemark{2}$\pm$ 0.01 & ... & ... & 1.57 $\pm$ 0.17 & ... & ... & ... & 19.43 $\pm$ 5.29 \\
\textbf{5123145(3)} & 3.77\tablenotemark{2}$\pm$ 0.01 & ... & ... & 5.91 $\pm$ 0.17 & ... & ... & ... & 0.43 $\pm$ 0.12 \\
\textbf{5129882} & 2.33 $\pm$ 0.06 & 48 $\pm$ 2 & 5.24 $\pm$ 0.20 & 4.93 $\pm$ 0.17\tablenotemark{3} & 10.14 & K3III & 0.79 $\pm$ 0.01 & 1.05 $\pm$ 0.27 \\
%
\textbf{5717541} & 1.19 $\pm$ 0.06 & 258 $\pm$ 2 & 3.44 $\pm$ 0.20 & 3.76 $\pm$ 0.17 & 8.71 & K3III & 4.05 $\pm$ 0.07 & 3.05 $\pm$ 0.84 \\
\textbf{5986270} & 2.82 $\pm$ 0.06 & 216 $\pm$ 2 & 3.27 $\pm$ 0.19 & 2.74 $\pm$ 0.17 & 10.37 & K3III & 4.70 $\pm$ 0.08 & 7.42 $\pm$ 1.94 \\
\textbf{6233558} &  3.38 $\pm$  0.06 & 43 $\pm$ 2 & 3.97 $\pm$ 0.19 & 3.58 $\pm$ 0.17 & 15.43 & K6III & 2.53 $\pm$ 0.004 & 3.57 $\pm$ 0.96 \\
\textbf{6356581(1)} & 3.36 $\pm$ 0.06 & 337 $\pm$ 2 & 6.10 $\pm$ 0.19 & 6.29 $\pm$ 0.19 & 27.38 & K3III & 0.36 $\pm$ 0.01 & 0.30 $\pm$ 0.09 \\
\textbf{6356581(2)} & 3.80\tablenotemark{2}$\pm$ 0.01 & ... & ... & 6.80 $\pm$ 0.18 & ... & ... & ... & 0.19 $\pm$ 0.06 \\
\textbf{6863041} & 0.59 $\pm$ 0.06\tablenotemark{1} & 132 $\pm$ 3 & 2.40 $\pm$ 0.20 & 2.45 $\pm$ 0.17 & 7.86 & G6V & 9.92 $\pm$ 0.16 & 9.50 $\pm$ 2.50 \\
\textbf{7529180} & 2.44 $\pm$ 0.06 & 253 $\pm$ 2 & 6.52 $\pm$ 0.19 & 4.39 $\pm$ 0.16 & 7.17 & F4V & 0.25 $\pm$ 0.004 & 1.72 $\pm$ 0.43 \\
\textbf{7630743} & 3.48 $\pm$ 0.06 & 90 $\pm$ 2 & 4.80 $\pm$ 0.20 & 4.86 $\pm$ 0.17 & 6.31 & K4III & 1.19 $\pm$ 0.02 & 1.12 $\pm$ 0.29 \\
\textbf{7690843} & 0.23 $\pm$ 0.06\tablenotemark{1} & 6 $\pm$ 8 & 0.81 $\pm$ 0.21 & 2.51 $\pm$ 0.17 & $<$3 & K3III & 32.05 $\pm$ 0.43 & 9.19 $\pm$ 2.34 \\
\textbf{7801848} & 0.37 $\pm$ 0.06 & 325 $\pm$ 5 & 0.36 $\pm$ 0.22 & 0.17 $\pm$ 0.18 & 5.27 & G3V & 41.74 $\pm$ 0.49 & 46.10 $\pm$ 12.48 \\
\textbf{7901207} & 2.53 $\pm$ 0.06 & 201 $\pm$ 2 & 5.82 $\pm$ 0.19 & 6.05 $\pm$ 0.17 & 35.35 & K3III & 0.47 $\pm$ 0.01 & 0.38 $\pm$ 0.10 \\
\textbf{8542853} & 0.85 $\pm$ 0.06 & 297 $\pm$ 3 & 0.34 $\pm$ 0.21 & 0.120 $\pm$ 0.18 & 5.60 & G7V & 42.13 $\pm$ 0.48 & 47.37 $\pm$ 12.54\\
\textbf{8983847(1)} & 2.54 $\pm$ 0.06 & 296 $\pm$ 2 & 4.90 $\pm$ 0.20 & 4.82 $\pm$ 0.17\tablenotemark{3} & 7.29 & K3III & 1.08 $\pm$ 0.02 & 1.17 $\pm$ 0.30 \\
\textbf{8983847(2)} & 3.23 $\pm$ 0.06 & 238 $\pm$ 2 & 6.11 $\pm$ 0.26 & 6.47 $\pm$ 0.17\tablenotemark{3} & ... & ... & 0.36 $\pm$ 0.01 & 0.26 $\pm$ 0.07 \\
\textbf{9702369} & 3.25 $\pm$ 0.06 & 353 $\pm$ 2 & 6.59 $\pm$ 0.19 & 4.72 $\pm$ 0.16 & 7.83 & F7V & 0.23 $\pm$ 0.004 & 1.27 $\pm$ 0.31 \\
\textbf{9965715} & 1.05 $\pm$ 0.06 & 152 $\pm$ 2 & 2.88 $\pm$ 0.21 & 2.11 $\pm$ 0.18 & 6.69 & F7V & 6.57 $\pm$ 0.12 & 12.55 $\pm$ 3.47 \\
\textbf{10124866} & 1.38 $\pm$ 0.06 & 186 $\pm$ 2 & 0.19 $\pm$ 0.21 & 0.17 $\pm$ 0.20 & 6.07 & G2V & 45.58 $\pm$ 0.49 & 46.12 $\pm$ 14.26 \\
\textbf{10140513} & 0.21 $\pm$ 0.06\tablenotemark{1} & 76 $\pm$ 9 & 0.45 $\pm$ 0.22 & 0.43 $\pm$ 0.16 & $<$3 & F9V & 39.78 $\pm$ 0.48 & 40.24 $\pm$ 9.57 \\
\textbf{10779537} & 1.93 $\pm$ 0.06 & 350 $\pm$ 2 & 4.54 $\pm$ 0.19 & 4.29 $\pm$ 0.17\tablenotemark{3} & 10.23 & K3III & 1.50 $\pm$ 0.026 & 1.88 $\pm$ 0.50 \\
\textbf{10797849} & 2.98 $\pm$ 0.06 & 271 $\pm$ 2 & 1.78 $\pm$ 0.20 & 1.63 $\pm$ 0.15 & 5.73 & F7V & 16.27 $\pm$ 0.25 & 18.24 $\pm$ 4.13 \\
\textbf{10909629} & 1.53 $\pm$ 0.06 & 59 $\pm$ 2 & 4.24 $\pm$ 0.20 & 3.88 $\pm$ 0.16\tablenotemark{3} & 10.41 & F7V & 1.98 $\pm$ 0.04 & 2.74 $\pm$ 0.68 \\
\textbf{11395018} & 3.25 $\pm$ 0.06 & 358 $\pm$ 2 & 2.95 $\pm$ 0.19 & 3.08 $\pm$ 0.17 & 8.21 & G3V & 6.17 $\pm$ 0.10 & 5.54 $\pm$ 1.43 \\
\textbf{11551430} & 3.60 $\pm$ 0.06 & 149 $\pm$ 2 & 2.02 $\pm$ 0.19 & 2.02 $\pm$ 0.15 & 12.98 & G5V & 13.51 $\pm$ 0.21 & 13.48 $\pm$ 3.09 \\

\hline
\end{longtable}

\tablenotetext{1}{Stars for which separation and position angle were determined with a PSF subtracted image.}
\tablenotetext{2}{Found in Keck image, not apparent in full or PSF Robo-AO image.}
\tablenotetext{3}{J-band image. KIC4999260 was taken in PK-continuum.}

\clearpage
\begin{longtable}{l || c  c c c c c c c c}
\caption{Individual Star Information for Companion Systems \label{CompDet3}}\\

\hline\\

\textbf{KIC ID} & \textbf{i\textquotesingle{}} & \textbf{K'} & \textbf{Spectral}  &  \textbf{Radius} & \textbf{Distance} & \textbf{Physical} & \textbf{Physical} & \textbf{Method} \\

\textbf{} & \textbf{Mag} & \textbf{Mag} & \textbf{Type} & \textbf{(R$_\odot$)} & \textbf{(pc)} & \textbf{Sep} & \textbf{Assoc} & \textbf{Used} \\

\textbf{} & \textbf{} & \textbf{} & \textbf{}  &  & & \textbf{[AU]} & \textbf{($\sigma$)\tablenotemark{1}} &\\

\textbf{1571152(A)} & 9.63 $\pm$ 0.11 & 8.82 $\pm$ 0.11 & F2V & \tablenotemark{*}$1.6^{+0.3}_{-0.2}$ & \tablenotemark{*}$321^{+51}_{-88}$ & ... & 0.57 & ...\\ 
\textbf{1571152(B)} & 10.54 $\pm$ 0.17 & 9.58 $\pm$ 0.14 & F7V & $1.5^{+0.3}_{-0.2}$ & $387^{+201}_{-111}$ & 128 & YES & A17\\
\textbf{}

\textbf{1576249(A)} & 11.82 $\pm$ 0.12 & 10.70 $\pm$ 0.68 & F7V & \tablenotemark{*}$1.6^\pm0.4$ & \tablenotemark{*}$381^{+82}_{-101}$ & ... & 0.29 & ...\\
\textbf{1576249(B)} & 12.44 $\pm$ 0.18 & 11.10 $\pm$ 1.78 & G7V & 1.2$\pm$0.2 & $447^{+221}_{-151}$ & 107 & YES & A17\\
\textbf{}

\textbf{1725815(A)} & 10.74 $\pm$ 0.10 & 9.64 $\pm$ 0.10 & F7V & $2.2^\pm0.1$ & $408\pm5$ & ... & 0.94 & ...\\
\textbf{1725815(B)} & 14.55 $\pm$ 0.17 & 12.86 $\pm$ 0.13 & K2V\tablenotemark{G} & 0.5 $\pm$ 0.1\tablenotemark{G} & 414 $\pm$ 4\tablenotemark{G} & 1486 & YES & Gaia\\

\textbf{2568519(A)} & 11.49 $\pm$ 0.60 & 10.72 $\pm$ 0.12 & F7V & \tablenotemark{*}$2.7^{+0.7}_{-0.9}$ & \tablenotemark{*}$584^{+144}_{-181}$ & ... & 0.11 & ...\\
\textbf{2568519(B)} & 12.53 $\pm$ 2.09 & 10.83 $\pm$ 0.17 & K1V & $1.2^{+0.9}_{-0.7}$ & $515^{+618}_{-368}$ & 93 & YES & A17\\

\textbf{3123191(A)} & 9.88 $\pm$ 0.11 & 8.81 $\pm$ 0.10 & F7V & $1.6\pm0.1$ & $196\pm4$ & ... & 1.47 & ...\\
\textbf{3123191(B)} & 11.78 $\pm$ 0.18 & 10.68 $\pm$ 0.12 & F2V & $1.3\pm0.2$ & $456^{+186}_{-129}$ & ... & NO & A17\\

\textbf{3221671(A)} & 8.99 $\pm$ 0.10 & 8.01 $\pm$ 0.11 & F5V & $1.6\pm0.1$ & $137^{+1}_{-0.4}$ & ... & 0.13/17.17 & ...\\
\textbf{3221671(B)} & 11.55 $\pm$ 0.17 & 9.48 $\pm$ 0.13 & K4III\tablenotemark{G} & 0.9$^{+0.04}_{-0.1}$ & 137 $\pm$ 1 & 222 & YES & Gaia\\
\textbf{3221671(C)} & ... & 12.21 $\pm$ 0.13 & ... & ... & 1820 $\pm$ 981 & ... & NO & Gaia\\

\textbf{3430893(A)} & 10.72 $\pm$ 0.10 & 9.55 $\pm$ 0.10 & F7V & $1.6^\pm0.1$ & $289\pm2$ & ... & ... & ...\\
\textbf{3430893(B)} & 14.09 $\pm$ 0.18 & 13.43 $\pm$ 0.13 & ... & ... & ... & ... & NO & Gaia\\

\textbf{3643774(A)} & 9.73 $\pm$ 0.10 & 8.58 $\pm$ 0.10 & G1V & $1.6\pm0.1$ & $187\pm1$ & ... & 6.30 & ...\\
\textbf{3643774(B)} & 14.87 $\pm$ 0.16 & 13.62 $\pm$ 0.13 & ... & ... & 760 $\pm$ 91\tablenotemark{G} & ... & NO & Gaia\\

\textbf{4260884(A)} & 11.06 $\pm$ 0.11 & 9.05 $\pm$ 0.11 & K3III & \tablenotemark{*}$13.2^{+0.9}_{-3.5}$ & \tablenotemark{*}$1439^{+122}_{-397}$ & ... & 3.11 & ...\\
\textbf{4260884(B)} & 12.12 $\pm$ 0.17 & 10.38 $\pm$ 0.13 & K2V & $1.0\pm0.1$ & $200^{+82}_{-35}$ & ... & NO & A17\\

\textbf{4914234(A)} & 11.40 $\pm$ 0.10 & 9.40 $\pm$ 0.10 & K3III & $11.8^{+0.7}_{-0.6}$ & $1721^{+73}_{-64}$ & ... & 0.24 & ...\\
\textbf{4914234(B)} & 17.67 $\pm$ 0.19 & 15.21 $\pm$ 0.12 & K8V & $0.8\pm0.1$ & 1586 $\pm$ 554\tablenotemark{G} & 6626 & YES & Gaia\\

\textbf{4999260(A)} & 9.17 $\pm$ 0.12 & 7.43 $\pm$ 0.10 & K3III & $1.3\pm0.1$ & $1320\pm43$ & ... & ... & ...\\
\textbf{4999260(B)} & 11.50 $\pm$ 0.18 & 11.38 $\pm$ 0.12 & ... & ... & ... & 1307 & YES & $<$1\farcs0\\

\textbf{5123145(A)} & 11.63 $\pm$ 0.10 & 9.75 $\pm$ 0.10 & K3III & \tablenotemark{*}$9.3^{+2.5}_{-3.8}$ & \tablenotemark{*}$1597^{+314}_{-681}$ & ... & 0.88/1.69/2.77&...\\
\textbf{5123145(B)} & 15.66 $\pm$ 0.17 & 13.66 $\pm$ 0.13 & K5V & $1.0^\pm0.1$ & $931^{+322}_{-147}$ & 4216 & YES & A17\\
\textbf{5123145(C)} & ... & 11.50 $\pm$ 0.14 & ... & ... & 2202 $\pm$ 171 & ... & NO & Gaia\\
\textbf{5123145(D)} & ... & 15.63 $\pm$ 0.14 & ... & ... & 227 $\pm$ 157 & ... & NO & Gaia\\

\textbf{5129882(A)} & 12.71 $\pm$ 0.10 & 10.87 $\pm$ 0.10 & K3III & $6.8^{+0.4}_{-0.3}$ & $2032^{+66}_{-58}$ & ... & 1.93 &...\\
\textbf{5129882(B)} & 17.95 $\pm$ 0.17 & 15.80 $\pm$ 0.13 & F6V & $1.1^{+0.2}_{-0.1}$ & $3775^{+1778}_{-902}$ & ... & NO & A17\\

\textbf{5717541(A)} & 11.65 $\pm$ 0.10 & 9.81 $\pm$ 0.10 & K3III & $5.6\pm0.3$ & $1000^{+23}_{-21}$ & ... & 1.01 &...\\
\textbf{5717541(B)} & 15.09 $\pm$ 0.17 & 13.56 $\pm$ 0.14 & ... & ... & $1101^{+520}_{-281}$ & 1190 & YES & A17\\

\textbf{5986270(A)} & 11.89 $\pm$ 0.10 & 9.74 $\pm$ 0.10 & K3III & $13.0\pm1.1$ & $2129^{+163}_{-127}$ & ... & 12.7 &...\\
\textbf{5986270(B)} & 15.15 $\pm$ 0.16 & 12.48 $\pm$ 0.13 & M1V & $0.8\pm0.1$ & $328^{+56}_{-53}$ & ... & NO & A17\\

\textbf{6233558(A)} & 12.27 $\pm$ 0.10 & 10.25 $\pm$ 0.10 & K6III & $6.0\pm0.3$ & $1208^{+41}_{-36}$ & ... & 5.18 &...\\
\textbf{6233558(B)} & 16.24 $\pm$ 0.16 & 13.83 $\pm$ 0.14 & K3V & $0.6\pm0.1$\tablenotemark{G} & 918 $\pm$ 43\tablenotemark{G} & ... & NO & Gaia\\

\textbf{6356581(A)} & 10.55 $\pm$ 0.10 & 8.87 $\pm$ 0.10 & K3III & $7.9\pm0.6$ & $906^{+23}_{-20}$ & ... & 2.31/...&...\\
\textbf{6356581(B)} & 16.65 $\pm$ 0.16 & 15.08 $\pm$ 0.15 & F3V & $1.1\pm0.1$ & $2681^{+1084}_{-768}$ & ... & NO & A17\\
\textbf{6356581(C)} & ...   & 15.80 $\pm$ 0.15 & ... & ... & ... & ... & NO & A17\\

\textbf{6863041(A)} & 11.37 $\pm$ 0.11 & 10.01 $\pm$ 0.10 & G6V & \tablenotemark{*}$2.4\pm0.1$ & \tablenotemark{*}$472^{+14}_{-24}$ & ... & 5.95 & ...\\
\textbf{6863041(B)} & 13.77 $\pm$ 0.17 & 12.45 $\pm$ 0.13 & ... & ... & $741^{+315}_{-211}$ & 279 & YES & A17\\

\textbf{7529180(A)} & 8.38 $\pm$ 0.10 & 7.47 $\pm$ 0.10 & F4V & $1.5\pm0.1$ & $110\pm1$ & ... & 0.83 & ...\\
\textbf{7529180(B)} & 14.91 $\pm$ 0.16 & 11.86 $\pm$ 0.13 & M3V & $0.4^{+0.2}_{-0.1}$ & $138^{+57}_{-39}$ & 269 & YES & A17\\

\textbf{7630743(A)} & 12.27 $\pm$ 0.10 & 10.31 $\pm$ 0.10 & K4III & $4.7\pm0.2$ & $1034^{+29}_{-25}$ & ... & 1.95 & ...\\
\textbf{7630743(B)} & 17.07 $\pm$ 0.17 & 15.17 $\pm$ 0.13 & K2III\tablenotemark{G} & ... & 4014 $\pm$ 1530\tablenotemark{G} & ... & NO & Gaia\\

\textbf{7690843(A)} & 11.18 $\pm$ 0.11 & 8.94 $\pm$ 0.10 & K3III & \tablenotemark{*}$10.5^{+3.1}_{-3.5}$ & \tablenotemark{*}$1161^{+365}_{-406}$ & ... & 0.38 &...\\
\textbf{7690843(B)} & 11.99 $\pm$ 0.21 & 11.45 $\pm$ 0.13 & ... & ... & $1478^{+325}_{-518}$ & 267 & YES & A17\\

\textbf{7801848(A)} & 9.77 $\pm$ 0.12 & 8.61 $\pm$ 0.11 & G3V & \tablenotemark{*}$0.9^{+0.2}_{-0.1}$ & \tablenotemark{*}$73^{+17}_{-7}$ & ... & 1.47 &...\\
\textbf{7801848(B)} & 10.13 $\pm$ 0.18 & 8.77 $\pm$ 0.14 & G5V & $1.1^{+0.2}_{-0.1}$ & $125^{+60}_{-33}$ & ... & NO & A17\\

\textbf{7901207(A)} & 11.02 $\pm$ 0.10 & 9.09 $\pm$ 0.10 & K3III & $10.9^{+0.6}_{-0.5}$ & $1422\pm51$ & ... & 0.84 & ...\\
\textbf{7901207(B)} & 16.84 $\pm$ 0.16 & 15.13 $\pm$ 0.13 & K3V & $1.1^{+0.5}_{-0.3}$ & 1485 $\pm$ 54\tablenotemark{G} & 3599 & YES & Gaia\\

\textbf{8542853(A)} & 9.74 $\pm$ 0.12 & 8.46 $\pm$ 0.11 & G7V & $1.3\pm0.1$ & $98^{+8}_{-6}$ & ... & 0.51 &...\\
\textbf{8542853(B)} & 10.08 $\pm$ 0.18 & 8.66 $\pm$ 0.14 & G7V & $1.1^{+0.2}_{-0.1}$ & 105 $\pm$ 10\tablenotemark{G} & 84 & YES & Gaia\\

\textbf{8983847(A)} & 12.87 $\pm$ 0.10 & 10.72 $\pm$ 0.10 & K3III & $4.3\pm0.2$ & $1144^{+23}_{-20}$ & ... & ... &...\\
\textbf{8983847(B)} & 17.78 $\pm$ 0.17 & 15.53 $\pm$ 0.13 & ... & ... & ... & ... & NO & A17\\
\textbf{8983847(C)} & 18.98 $\pm$ 0.23 & 17.18 $\pm$ 0.13 & ... & ... & ... & ... & NO & A17\\

\textbf{9702369(A)} & 9.45 $\pm$ 0.10 & 8.43 $\pm$ 0.10 & F7V & $1.3\pm0.1$ & $144\pm2$ & ... & 2.13 &...\\
\textbf{9702369(B)} & 16.04 $\pm$ 0.16 & 13.16 $\pm$ 0.12 & M2V & 0.5 $\pm$ 0.1 & $307^{+87}_{-89}$ & ... & NO & A17\\

\textbf{9965715(A)} & 9.33 $\pm$ 0.10 & 8.02 $\pm$ 0.10 & F7V & $1.4\pm0.1$ & $124\pm0.4$ & ... & 0.32 &...\\
\textbf{9965715(B)} & 12.21 $\pm$ 0.18 & 10.13 $\pm$ 0.14 & K5V & $0.9^{+0.04}_{-0.1}$ & $132^{+22}_{-20}$ & 130 & YES & A17\\

\textbf{10124866(A)} & 8.46 $\pm$ 0.12 & 7.01 $\pm$ 0.12 & G2V & $1.3\pm0.1$ & $53^\pm0.1$ & ... & 2.42 &...\\
\textbf{10124866(B)} & 8.66 $\pm$ 0.18 & 7.18 $\pm$ 0.16 & G8V & $1.0^{+0.2}_{-0.1}$ & 52 $\pm$ 0.2\tablenotemark{G} & 73 & YES & Gaia\\

\textbf{10140513(A)} & 11.42 $\pm$ 0.12 & 10.20 $\pm$ 0.11 & F9V & \tablenotemark{*}$1.1^\pm0.2$ & \tablenotemark{*}$203^{+28}_{-34}$ & ... & 1.47 &...\\
\textbf{10140513(B)} & 11.87 $\pm$ 0.18 & 10.63 $\pm$ 0.12 & F5V & $1.2\pm0.2$ & $379^{+157}_{-117}$ & ... & NO & A17\\

\textbf{10779537(A)} & 11.80 $\pm$ 0.10 & 9.95 $\pm$ 0.10 & K3III & $4.7\pm0.2$ & $900^{+20}_{-19}$ & ... & ... &...\\
\textbf{10779537(B)} & 16.35 $\pm$ 0.19 & 14.24 $\pm$ 0.13 & ... & ... & ... & ... & NO & A17\\

\textbf{10797849(A)} & 10.90 $\pm$ 0.11 & 9.68 $\pm$ 0.10 & F7V & $2.3\pm0.2$ & $414^\pm4$ & ... & 1.03 & ...\\
\textbf{10797849(B)} & 12.68 $\pm$ 0.16 & 11.30 $\pm$ 0.11 & G5V & $1.1^{+0.2}_{-0.1}$\tablenotemark{G} & 423 $\pm$ 5\tablenotemark{G} & 1233 & YES & Gaia\\

\textbf{10909629(A)} & 10.79 $\pm$ 0.10 & 9.65 $\pm$ 0.10 & F7V & $2.2\pm0.1$ & $429^{+5}_{-4}$ & ... & 2.03 &...\\
\textbf{10909629(B)} & 15.03 $\pm$ 0.17 & 13.12 $\pm$ 0.13 & ... & ... & 515 $\pm$ 42 & ... & NO & Gaia\\

\textbf{11395018(A)} & 10.63 $\pm$ 0.10 & 9.28 $\pm$ 0.10 & G3V & $2.2\pm0.1$ & $340^{+3}_{-2}$ & ... & 30.70 &...\\
\textbf{11395018(B)} & 13.59 $\pm$ 0.16 & 12.36 $\pm$ 0.13 & F5V & $1.3\pm0.2$ & $913^{+355}_{-259}$ & ... & NO & Gaia\\

\textbf{11551430(A)} & 10.62 $\pm$ 0.11 & 9.15 $\pm$ 0.10 & G5V & $2.4\pm0.1$ & $324^\pm4$ & ... & 5.19 &...\\
\textbf{11551430(B)} & 12.64 $\pm$ 0.16 & 11.17 $\pm$ 0.11 & G7V & $1.1^{+0.2}_{-0.1}$\tablenotemark{G} & 348 $\pm$ 3\tablenotemark{G} & 1166 & YES & Gaia\\

\end{longtable}

\tablenotetext{*}{Value taken from the KSPC. If not marked, values come from \cite{Berger2018}.}
\tablenotetext{G}{Value calculated from values on Gaia DR2 Database.}
\tablenotetext{1}{Likelihood of star \textit{not} being physically associated.}

\pagebreak

\appendix

\section{Companion Figure}
\label{fig:appendix1}

\begin{figure}[ht]
\centering
\includegraphics[scale = 0.125]{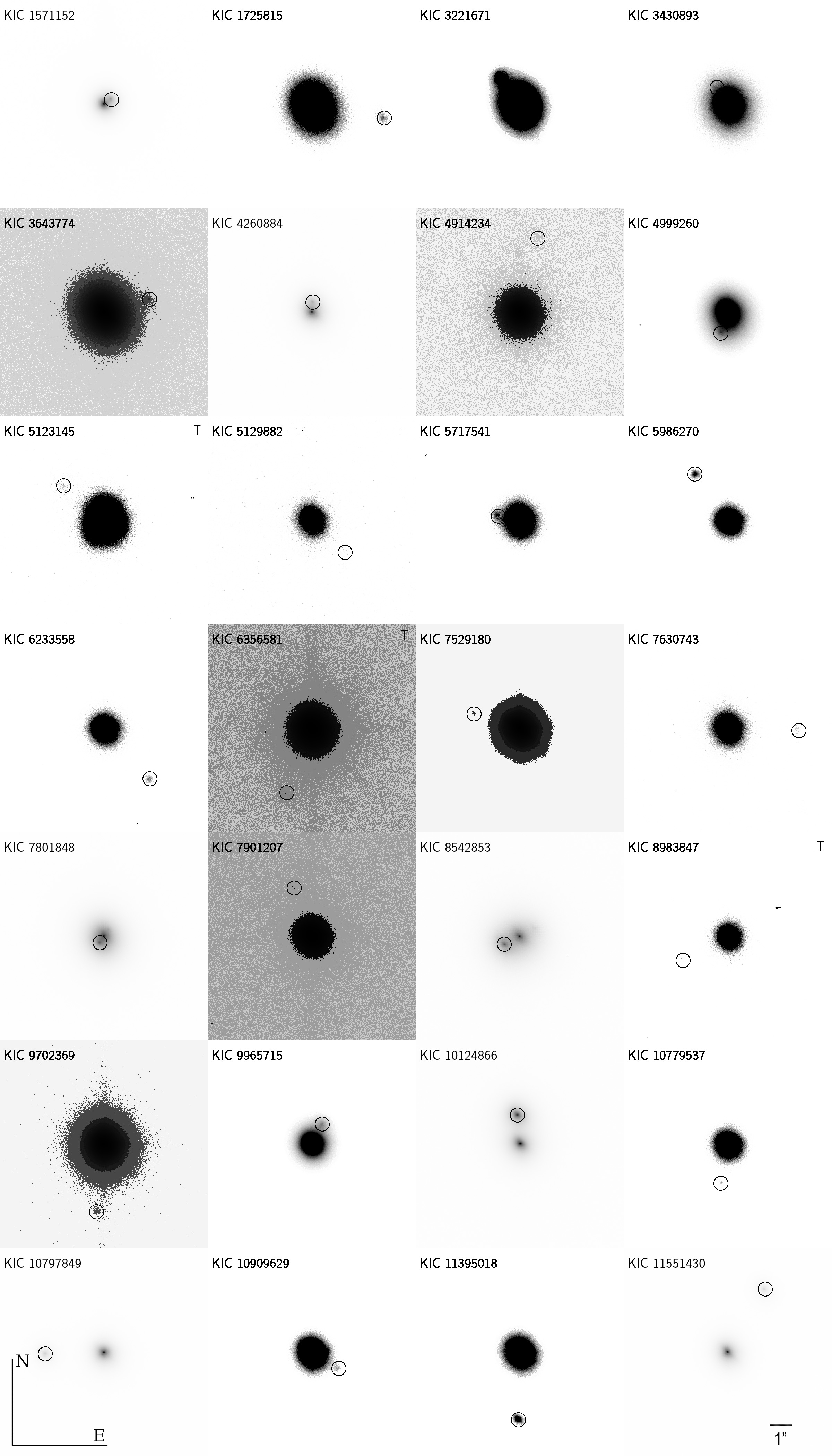}
\caption{Robo-AO \textit{i}\textquotesingle{}-band images of candidate companion systems. Images have been centered and cropped to the primary target and circles drawn around the companion. For companions that were not directly visible, Python module image enhancers Pillow and cv2 have been used to alter the contrast until the secondary is visible. Inverting of images was performed with Pillow. Images with T in the top right corner are triple systems but only the Robo-AO discoveries are circled.}
\label{fig:Discoveries}
\end{figure}

\section{PSF Companion Figure}
\label{fig:appendix2}
\begin{figure}[ht]
\centering
\includegraphics[scale = 1]{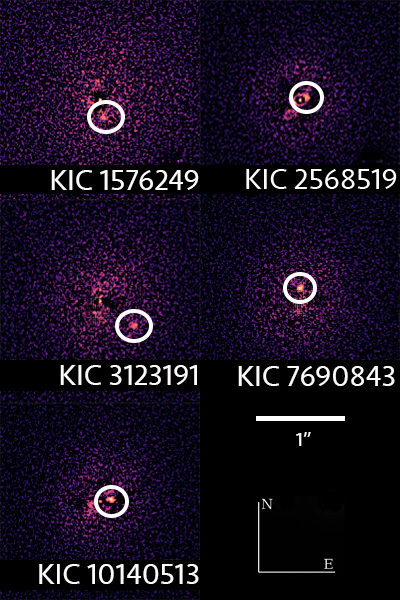}
\caption{Robo-AO \textit{i}\textquotesingle{}-band PSF-subtracted images of discovered candidate companion systems.}
\label{fig:Discoveries_PSF}
\end{figure}

\end{document}